\documentclass[aps,prd,twocolumn,amsmath,amssymb,showpacs,floatfix,nofootinbib]{revtex4-1}

\usepackage{amsmath}
\usepackage{hyperref}
\usepackage{graphicx}
\usepackage{xcolor}
\usepackage{hhline}
\usepackage{cancel}
\usepackage[normalem]{ulem}
\usepackage{amssymb}
\usepackage{amsfonts}
\usepackage{enumerate}
\usepackage{afterpage}
\usepackage{color}
\usepackage{ae,aecompl}
\usepackage{soul}

\newcommand{\be}{\begin{equation}}
\newcommand{\ee}{\end{equation}}

\begin{document}

\title[The nature of progenitors of BH mergers]{Determining the progenitors of merging black-hole binaries}
\author{Alvise Raccanelli}
\author{Ely D. Kovetz}
\author{Simeon Bird}
\author{Ilias Cholis}
\author{Julian B. Mu\~{n}oz}
\affiliation{%
 Department of Physics \& Astronomy, Johns Hopkins University, 3400 N. Charles 
St., Baltimore, MD 21218, USA \\
}%

\begin{abstract}
We investigate a possible method for determining the progenitors of black hole (BH) mergers 
observed via their gravitational wave (GW) signal.
We argue that measurements of the cross-correlation of the GW events with 
overlapping galaxy catalogs may provide an additional tool in determining if BH mergers trace the stellar 
mass of the Universe, as would be expected from mergers of the endpoints of stellar evolution.
If on the other hand the BHs are of primordial origin, as has been recently suggested, 
their merging would be preferentially hosted by lower biased objects, and thus have a lower 
cross-correlation with luminous galaxies.
Here we forecast the expected precision of the cross-correlation measurement for current and future GW 
detectors such as LIGO and the Einstein Telescope. We then predict how well these instruments 
can distinguish the model that identifies high-mass BH-BH mergers as 
the merger of primordial black holes that constitute the dark matter in the Universe from more traditional astrophysical sources.
\end{abstract}

\maketitle

\section{Introduction}
The recent detection of gravitational waves (GWs) from the merger of two black 
holes (BHs) of mass $\sim 30\, M_\odot$ by the LIGO 
collaboration~\cite{LIGO:GW} has confirmed the existence of GWs and 
opened up a new era of GW astronomy.
However, the nature of the progenitors of this high-mass BH-binary remains in question.

The fact that the first GW-signal detected was from a pair of relatively 
high-mass merging BHs suggests that such events are 
common enough that a significant sample of them will soon be obtained. 
However, since these BH-BH mergers are not generically expected to 
be accompanied by electromagnetic (EM) counterparts (see, though,
~\cite{Connaughton:2016umz, Loeb:2016fzn, Kotera:2016dmp}), their localization to specific host 
galaxies is most likely impossible. 

In this work we study the progenitor question statistically, via the 
cross-correlation of GW events with galaxy catalogs. The amplitude 
of the cross-correlation depends on the bias, 
redshift distribution and clustering properties of the GW host halos.
For example, GW events produced by merging BHs inside globular 
clusters~\cite{Chatterjee:2016}, as an endpoint of stellar evolution in 
galaxies, are expected to roughly trace the stellar mass content of the Universe.
In this case GW and galaxy catalogs would be highly correlated. 
However, in alternative models whereby BH binaries reside mostly within halos of 
particular masses, or exhibit different redshift and angular distributions, 
the cross-correlation with galaxies would be weaker.

The possibility of correlating GWs with galaxies in order to 
determine if BH-binaries trace the matter inhomogeneities in the Universe 
has been investigated in~\cite{Namikawa:2016}.
Our analysis uses similar tools, but extends them towards a novel goal: using the 
cross-correlation as a method to probe the nature of BH binary progenitors. 

One alternative hypothesis for the BH merger progenitors is 
primordial black holes (PBHs) which could make up the dark matter in the 
Universe~\cite{Bird:2016} (see also~\cite{Garcia-Bellido:1996, Nakamura:1997, Clesse:2016, Sasaki:2016}).
In this scenario, PBH-PBH mergers occur preferentially in low-mass 
halos, which are more uniformly distributed, and are a less biased  
tracer of the dark-matter distribution than star-forming galaxies. 
We investigate how well current and future instruments could use 
measurements of the cross-correlation between BH mergers and 
luminous galaxies to test this model. More generally, these GW maps 
can be cross-correlated with catalogs of alternative galaxy 
populations, with different biases and redshift distributions, 
to test a wider family of potential BH-binary progenitor models~\cite{Kinugawa:2014zha, Inayoshi:2016hco, Hartwig:2016nde}. 
This method can also be extended to other types of GW signals, such as those
originating from tidal disruption events by super-massive black holes~\cite{OLeary:2008, Stone:2012uk, Ali-Haimoud:2015bfg}, 
or those coming from neutron-star binaries, once these observations can 
reach a cosmological volume, as will be possible for upcoming GW experiments~\cite{Sathyaprakash:2011}.

The structure of this paper is as follows. In Section~\ref{sec:corr} 
we describe our methods for measuring the cross-correlation 
of GW sources with other structure tracers, including the GW and galaxy catalogs we consider. 
In Section~\ref{sec:Results} we present the results we forecast for a general BH 
population, followed by those for the PBH scenario. We then summarize our findings and conclude in 
Section~\ref{sec:conclusions}.


\section{Methods}
\label{sec:corr}

\subsection{Galaxy and GW correlations}
\label{sec:crosscorr}
In order to measure the correlation between the host halos of BH-binaries 
and galaxies, we use measurements of their number counts.
We consider angular projections $C_\ell$, that can be calculated from 
the underlying 3D matter power spectrum by using 
(see e.g.~\cite{Raccanelli:2008, Pullen:2012}):
\begin{equation}
\label{eq:ClXY}
C_{\ell}^{XY} = r \int \frac{4\pi dk}{k} \Delta^2(k) W_{\ell}^X(k) W_{\ell}^Y(k) \, ,
\end{equation}
where $W_{\ell}^{\{X,Y\}}$ are the source distribution window functions 
for the different observables (here $X$ and $Y$ stand for galaxies and GWs), $\Delta^2(k)$ 
is the dimensionless matter power spectrum today, and $r$ is a 
cross-correlation coefficient ($r\equiv1$ for the auto-correlation case, $X=Y$).

The window function for the number count distributions can be written as 
(see e.g.~\cite{Cabre:2007}):
\begin{equation}
\label{eq:flg}
W_{\ell}^X(k) = \int \frac{\bar{d N_X}(z)}{dz} b_X(z) j_{\ell}[k\chi(z)] dz \, .
\end{equation}
$d \bar{N}_X(z)/dz$ is the source redshift distribution, normalized to 
unity within the same redshift range as the window function;
$b_X(z)$ is the bias that relates the observed correlation function to the 
underlying matter distribution, that we assume to be scale-independent on large scales;
$j_{\ell}(x)$ is the spherical Bessel function of order $\ell$, and $\chi(z)$ is the comoving distance.
The integral in Equation~\eqref{eq:flg} is performed over the redshift range corresponding to the selection 
function of the galaxy survey.

As explained in Section \ref{sec:catalogs}, for our galaxy catalog we assume 
a constant redshift distribution of galaxies. As for GW events, their number can be estimated by:
\begin{equation}
\label{eq:ngw}
\frac{d N_{GW}(z)}{dz} \approx \mathcal{R}(z) T_{\rm obs} \frac{4\pi\chi^2(z)}{(1+z)H(z)} \, ,
\end{equation}
where $\mathcal{R}(z)$ is the redshift-dependent merger rate, 
$T_{\rm obs}$ is the observation time and $H(z)$ is the Hubble parameter.
The errors in the auto- and cross-correlations are given by 
(see e.g.~\cite{Cabre:2007, DiDio:2014}):
\begin{equation}
\label{eq:err-clgt}
\sigma_{C_{\ell}^{\rm g\,GW}} = \sqrt{\frac{\left(C_{\ell}^{\rm g\,GW}\right)^2 
+ \left[ \left( C_{\ell}^{gg} + \frac{1}{\bar{n}_g} \right) \left(C_{\ell}^{\rm 
GW \, GW}+ \frac{1}{\bar{n}_{\rm GW}} \right)\right]}{(2\ell+1)f_{\rm sky}}} \, ,
\end{equation}
and:
\begin{equation}
\label{eq:err-clgg}
\sigma_{C_{\ell}^{\rm g\,g}} = \sqrt{\frac{2 \left[C_{\ell}^{gg} + 
\frac{1}{\bar{n}_g}\right]^2}{(2\ell+1)f_{\rm sky}}} \, ,
\end{equation}
where $f_{\rm sky}$ is the fraction of the sky observed and 
$\bar{n}_{\{\rm g, GW\}}$ is the average number of sources per 
steradian, i.e. the integral of $dN/dz$ (Equation~\eqref{eq:ngw} in the GW case).

Our analysis takes into account the uncertainty in the value of the 
galaxy bias, by estimating the precision with which it can be measured 
using the galaxy auto-correlation power spectrum, which is then used 
as a prior in the Fisher analysis. 
Alternative probes using external datasets, such as galaxy---CMB-lensing 
correlations~\cite{Vallinotto:2013, Giannantonio:2014, Pujol:2016, Chang:2016}, 
can potentially provide more accurate priors. 
In Section~\ref{sec:Results} we investigate the issue of galaxy 
bias uncertainty in more detail. 

\subsection{Galaxy catalogs}
\label{sec:catalogs}
While the error on the cross-correlation between GW and galaxies is dominated by the 
number of GW events observed, to obtain quantitative estimates we must assume a fiducial galaxy catalog.
The quantities which enter into our analysis are the number density of galaxies used, the bias of the 
specific observed galaxy sample. Concretely, we assume $b_G = 1.4$, and a galaxy number density of $4000$ deg.$^{-2}$. 
The bias and the source redshift distribution $dN/dz$ are assumed to be constant with redshift.

These assumptions are similar to those predicted for a galaxy survey 
resembling the planned Square Kilometer Array (SKA) wide and deep radio
survey~\cite{SKA:Jarvis}, estimated using the prescription of~\cite{Wilman:2008}.
We emphasize, however, that the main bottleneck in 
determining the progenitors of GW events is the number of GW 
events detected, rather than the details of the galaxy survey used. 
When computing the cross-correlation, the galaxy bias drops out (if assumed constant in redshift), and, 
as we shall discuss in Section~\ref{sec:Results}, our results are insensitive 
to the number density of galaxies, provided it is above a sufficient level.

Our results will be computed assuming that approximate redshifts are available for 
the galaxy catalogue. In the case of optical surveys, 
redshift information would be readily available, while for radio continuum 
surveys, redshift-binning could be obtained by using methods such as 
clustering-based redshift estimation~\cite{Menard:2013, Rahman:2014, Kovetz:CBR}.

\subsection{GW Experiments}
\label{sec:GWexp}

We shall consider four different Earth-based GW detectors/data configurations. 
As all GW detectors are full-sky experiments, and earth-based experiments probe 
similar frequency ranges, we shall distinguish them by the sensitivity, in terms 
of the redshift range probed, and the minimum angular scale to which the GW events 
can be localized. The issue of spatial localization is complicated and has been investigated 
in detail, see e.g.~\cite{Schutz:2011, Klimenko:2011, Sidery:2014, Namikawa:2016noz}.
The exact value of $\ell_{\rm max}$ will depend in principle on redshift, 
position on the sky, SNR of the event, and a variety of instrument design 
parameters\footnote{There have been proposals and studies on the advantages 
of building multiple detectors in a variety of locations, see e.g.~\cite{LIGO-net}.}.
An accurate determination of this value for all events is beyond the scope of this paper, 
and so we use a constant value for the angular resolution.
We use the following specifications:
\begin{enumerate}[i)]
	\item aLIGO + VIRGO: $\ell_{\rm max} = 20$, $z_{\rm max} = 0.75$;
	\item LIGO-net: $\ell_{\rm max} = 50$, $z_{\rm max} = 1.0$;
	\item Einstein Telescope: $\ell_{\rm max} = 100$, $z_{\rm max} = 1.5$;
	\item Einstein Telescope binned: $\ell_{\rm max} = 100$, $z_{\rm max} = 1.5$, \\ binned with $0<z_1<0.75<z_2<1.5$.
\end{enumerate}
Here $\ell_{\rm max}$ (=180$^{\circ}/\theta$) is the multipole corresponding 
to the finest angular resolution $\theta$ at which the GW events can be localized 
and $z_{\rm max}$ is the maximum redshift to which each experiment can detect a GW event.

``Einstein Telescope binned'' shows results for the Einstein telescope GW catalog, divided into two redshift bins.
Redshift binning can increase our ability to cross-correlate the GW catalog with other sources at the expense 
of decreasing the number counts and thus increasing the shot noise. For smaller experiments the expected number of
events detected is small, and a division of its catalog into bins renders the shot noise term prohibitive.


\subsection{GW Merger rates}
As shown in Section~\ref{sec:crosscorr}, the error on the cross-correlation depends on the 
shot noise in the gravitational wave sources, proportional to the number of gravitational 
wave events, $\bar{n}_{\rm GW}$. We shall see that this term frequently dominates the total 
error. We shall parametrize $\bar{n}_{\rm GW}$ with the {\it integrated} merger rate $\mathcal{R}$. 
Increased merger rates will provide better constraining power, by reducing the GW shot noise. 
We emphasize that while our forecast constraints depend strongly on the observed merger rate, by
the time the measurement is to be made, the merger rate will be known extremely well.

The total merger rate for all BH-BH merger events implied by the current LIGO detection 
is 2-400 Gpc$^{-3}$yr$^{-1}$~\cite{LIGO:rate} for $z < 0.5$. Given the current large uncertainty, 
we adopt throughout a fiducial value of 50 Gpc$^{-3}$yr$^{-1}$ averaged over $z \leq 0.5$, 
and include predictions for a range from 30 to 100 Gpc$^{-3}$yr$^{-1}$. This matches the merger rate 
expected from BH mergers resulting as the end-point of stellar binary evolution 
from~\cite{Dominik:2013}. For the redshift evolution of the rate $R(z)$, 
following~\cite{Dominik:2013} we assume for simplicity that environments with a 
metallicity of $0.25 Z_{\odot}$ are the dominant contributor to BH-BH binary mergers.
Given that the formation process of BH binaries is currently highly uncertain, this assumption 
on the metallicity is a reasonable ansatz.

We also need an estimate for the merger rate from the $30 M_\odot$ PBHs we suggest may 
comprise the dark matter. Here we shall follow theoretical expectations from~\cite{Bird:2016}, 
which suggest that the merger rate is $\mathcal{R} \approx 3$ Gpc$^{-3}$yr$^{-1}$, constant with redshift.
However, this estimate includes several large and difficult to quantify theoretical uncertainties. 
To reflect this we will consider a range of merger rates between 1 and 6 Gpc$^{-3}$yr$^{-1}$.

Note that these two estimates are not exclusive; the total rate of BH mergers is independent 
of the rate of $30 M_\odot$ mergers from PBHs.

In principle, GW number counts are modified by gravitational lensing in two ways. First, by 
changing their apparent angular position due to lensing convergence. Secondly, 
their {\it observed} number density is changed due to cosmic magnification by the intervening 
mass distribution~\cite{Matsubara:2000, Camera:2013, Oguri:2016}. However, these effects are
important only on small scales, which ground-based GW detectors do not have access 
to (assuming there are no EM counterparts), so we shall safely neglect them.


\subsection{GW Bias}
\begin{figure}
\centering
\includegraphics[width=\columnwidth]{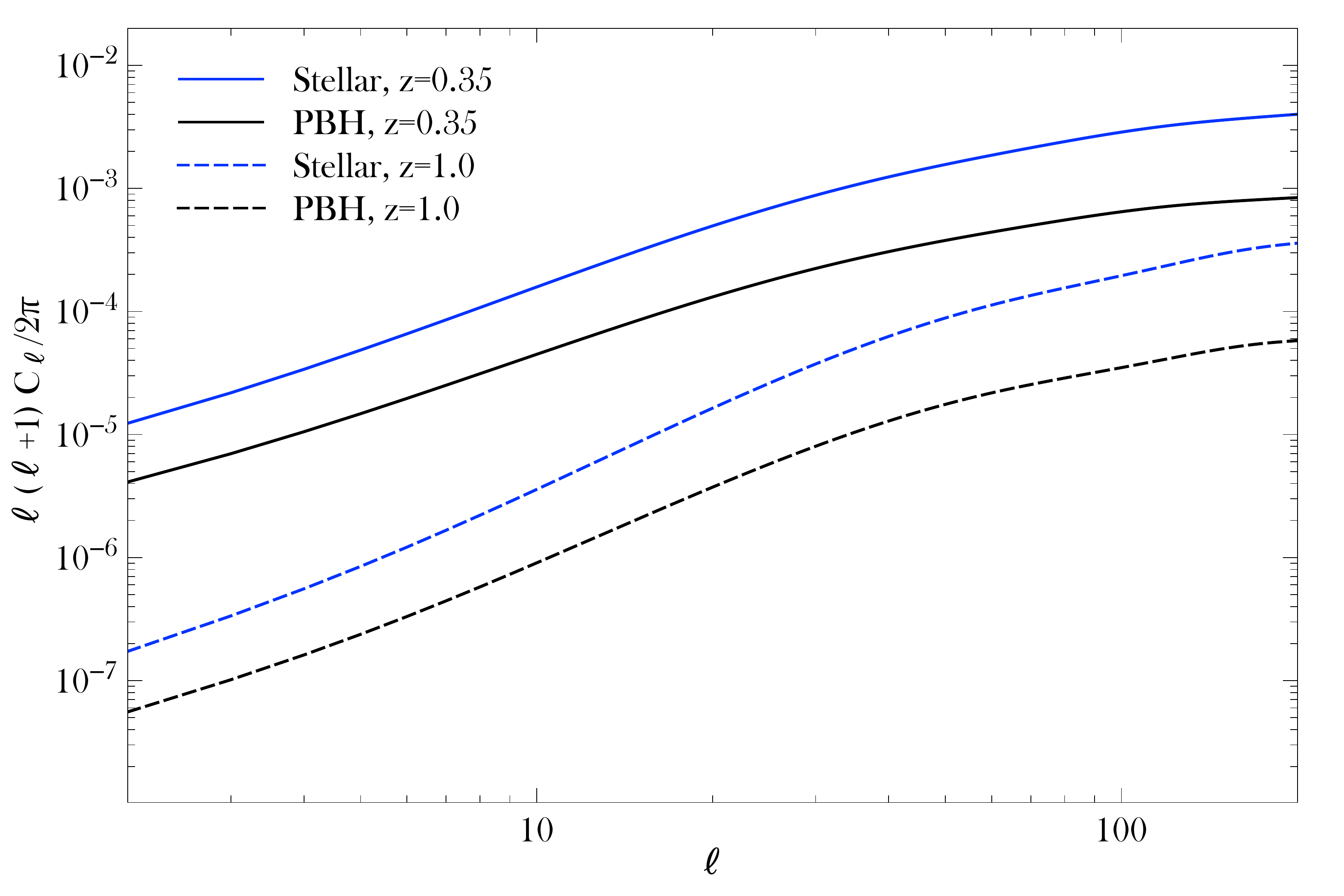}
\caption{Forecast amplitude of the cross-correlation between our fiducial galaxy sample and BH mergers as a function of multipole $\ell$. Solid lines 
show the results for $z=0.5$, and dashed lines for $z=1.0$, both integrated over a redshift shell of width $\Delta_z=0.35$. The two blue lines correspond 
to our fiducial model for BH mergers of stellar origin, in halos with $b_{GW}^{\rm Stellar} = 1.4$, while the two black lines correspond to 
mergers resulting from PBHs, with $b_{GW}^{\rm PBH} = 0.5$. We assume $r=1$ for both cases.}
\label{fig:corr_reg}
\end{figure}


As discussed above, our goal is to distinguish between different progenitor 
models by measuring the bias of the GW sources from the linear matter power spectrum.
GW events resulting from the endpoints of stellar binary evolution in a halo are expected 
to be a function of the star formation rate and the metallicity in the halo. They will thus tend 
to occur in larger and more heavily biased halos than mergers from PBHs, which 
have been shown to occur predominantly in small halos below the threshold for forming stars~\cite{Bird:2016}.
The bias for small halos can be estimated analytically using (see e.g.~\cite{Mo:1995}):
\begin{equation}
\label{eq:bh}
b_{\rm halo} = 1 +  \frac{\nu^2 -1}{\delta_c}\, ,
\end{equation}
where $\delta_c = 1.686$ is the critical overdensity value for spherical collapse,
and $\nu \equiv \delta_c/\sigma(M)$, where $\sigma(M)$ is the mass variance.
Equation~\eqref{eq:bh} gives $b_{\rm halo} \sim 0.45$ at $z=0$, 
and $b_{\rm halo} \sim 0.5$ at $z=1.5$ for $M < 10^6 M_\odot$. As this includes 
the overwhelming majority of halos hosting PBH mergers, we will take $b_{GW}^{\rm PBH} = 0.5$, 
constant with redshift.

For BH mergers with stellar binary progenitors, we assume the galaxies that host the majority of the stars
have similar properties to our observed galaxy sample. Thus we assume the same bias for stellar
GW binaries as we assumed for our galaxy sample in Section~\ref{sec:catalogs}, $b^{\rm Stellar}_{GW} = b_g = 1.4$.
We assume this bias is constant with redshift; in practice the bias of, for example, 
a $10^{12} M_\odot$ halo will be larger at higher redshift, as objects of that size become rarer. 
This will increase $\Delta b = b_{GW}^{\rm Stellar} - b_{GW}^{\rm PBH}$, making our estimates conservative.

Thus, if we cross-correlate a GW event map (filtered to contain only $\gtrsim 30\,M_\odot$ events) 
with a galaxy catalog, under the assumption that the progenitors of BH-binaries in this 
mass range are primarily dark matter PBHs, we would expect a bias difference of 
$\Delta b = b_{GW}^{\rm Stellar} - b_{GW}^{\rm PBH} \gtrsim 0.9$. If we instead 
assume that BH binaries form as the endpoint of stellar evolution, we expect
$\Delta b \sim 0$. In Figure~\ref{fig:corr_reg} we show the predicted cross-correlation of our 
galaxy catalog for both models; BH mergers of primordial and stellar origin.

\subsection{Estimating the cross-correlation amplitude}
\label{sec:estimator}
We now introduce a minimum-variance estimator for the {\it effective correlation amplitude}, 
$A_c\equiv r \times b_{GW}$, where $r$ is the cross-correlation coefficient of Equation~\eqref{eq:ClXY}. 
This cross-correlation coefficient parametrizes the extent to which two biased tracers of the matter field 
are correlated~\cite{Tegmark:1998}. In our case, since we are only 
interested in large angular scales, substantially larger than the size of the halos concerned, 
it is reasonable to expect that $r \approx 1$\footnote{The cross-correlation coefficient can 
be smaller than unity. For example, if in any dynamical process the binaries are ejected far 
away from their host galaxy, $r$ would reflect the fraction remaining in their hosts. This effect 
is not important unless very high angular-resolution is achievable.}. Nevertheless, in what 
follows we constrain $A_c$, for full generality.
 
The minimum-variance estimator for the effective correlation amplitude is given by (see e.g.~\cite{Jeong:2012fossils, Dai:2016}):
\begin{equation}
\label{eq:Ac}
\widehat{A_c} = \frac{\Sigma_\ell {\tilde{C}_\ell} F_\ell / {\rm Var}[\tilde{C}_\ell] 
}{\Sigma_\ell F_\ell^2 / {\rm Var}[\tilde{C}_\ell]} \, ,
\end{equation}
where $\tilde{C}_\ell$ is the measured power spectrum and $F_\ell \equiv{d \tilde{C}_\ell}/{d \widehat{A_c}}\propto b_g$. 
The variance of this estimator is then:
\begin{equation}
\sigma^2_{\widehat{A_c}} = \left[\sum_\ell \frac{F_\ell^2}{{\rm Var}[\tilde{C}_\ell]}\right]^{-1} \, ,
\end{equation}
which can be used to forecast the measurement error when neglecting that of other parameters.

More generally, the measurement error for specific parameters in a given experiment can 
be estimated using Fisher analysis. We write the Fisher matrix as:
\begin{equation}
\label{eq:Fisher}
F_{\alpha\beta} = \sum_{\ell} \frac{\partial C_\ell}{\partial 
\vartheta_\alpha}
\frac{\partial C_\ell}{\partial
\vartheta_\beta} {\sigma_{C_\ell}^{-2}} \, , 
\end{equation}
where $\vartheta_{\alpha} = \{ A_c, b_g \}$;
the derivatives of the power spectra $C_\ell$ are evaluated at fiducial values 
$\bar \vartheta_{\alpha}$ corresponding to the scenario at hand, 
and $\sigma_{C_\ell}$ are errors in the power spectra.

We obtain our results by computing the $2\times2$ Fisher matrix for the 
parameters $\{A_c, b_g\}$, using a prior on the galaxy bias corresponding 
to the precision reached by fitting the amplitude of the galaxy auto-correlation 
function $C_\ell^{gg}$. For this galaxy auto-correlation we can use a larger $\ell_{\rm max}$, 
because we are not limited by the poor spatial localization in the detection of 
GWs. We therefore use a value of $\ell_{\rm max}=200$, which yields a 
$\sim 10\%$ precision in the measurement of the bias $b_g$.
The impact of allowing the galaxy bias to vary in a wider 
range will be discussed in Section~\ref{sec:Results}. 

When using multiple redshift bins we neglect the 
correlation between different bins; as our assumed redshift bins are wide, 
and we do not include cosmic magnification, 
this cross-correlation contains virtually no information.
On large scales, galaxy clustering should be in principle modified to 
account for general-relativistic corrections (see e.g.~\cite{Yoo:2010, 
Bonvin:2011, Challinor:2011, Yoo:2012, Jeong:2012, Bertacca:2012, DiDio:2014}). 
However, these effects are subdominant compared to the 
uncertainty in the merger rate and we verified that our conclusions are not 
heavily affected by neglecting them (for a study on the impact of GR effects on 
cosmological parameter estimation, see~\cite{Raccanelli:2015GR}).


\section{Results}
\label{sec:Results}

\begin{figure*}[ht]
\centering
\includegraphics[width=0.49\textwidth]{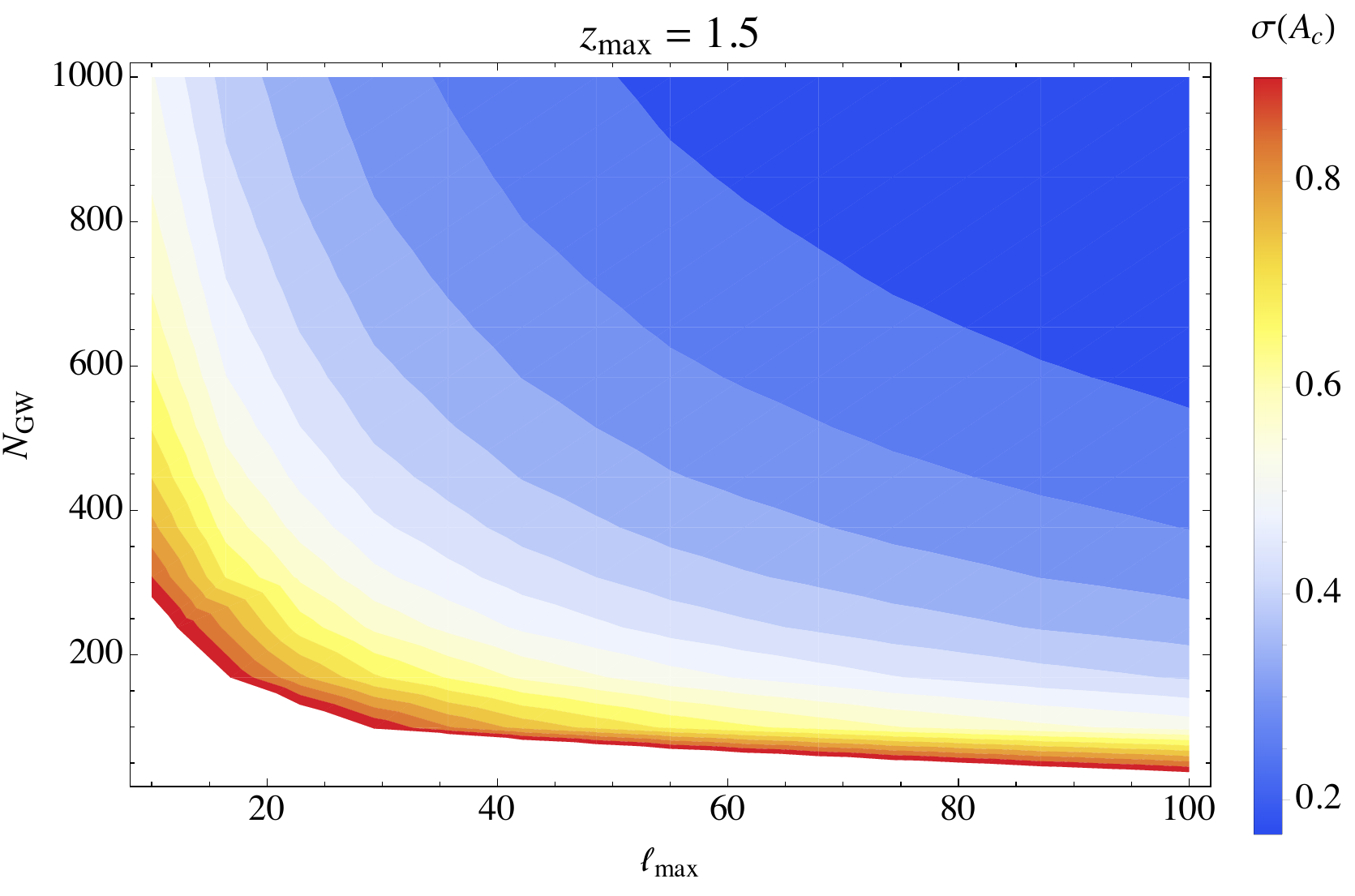}
\includegraphics[width=0.49\textwidth]{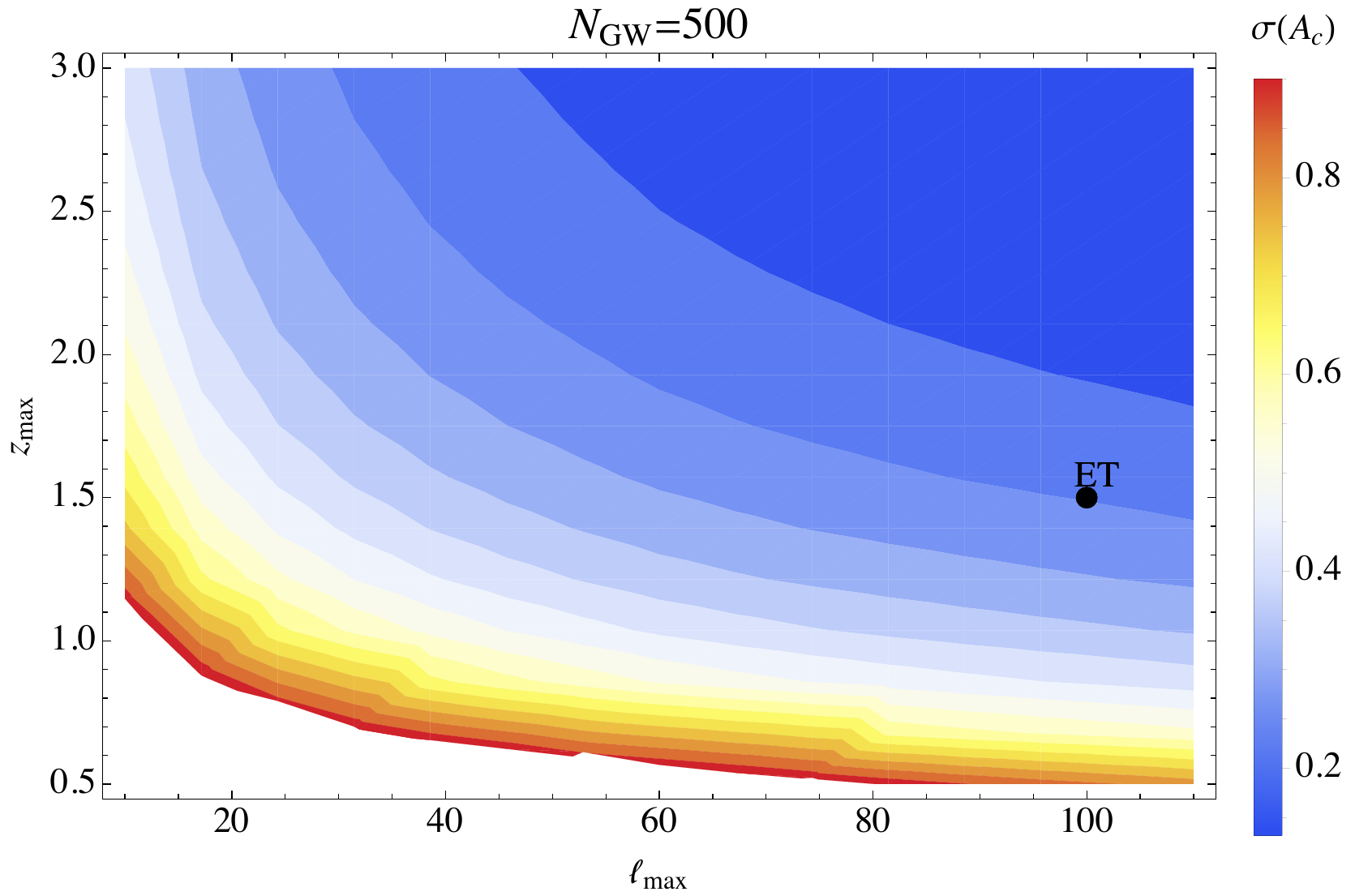}
\includegraphics[width=0.49\textwidth]{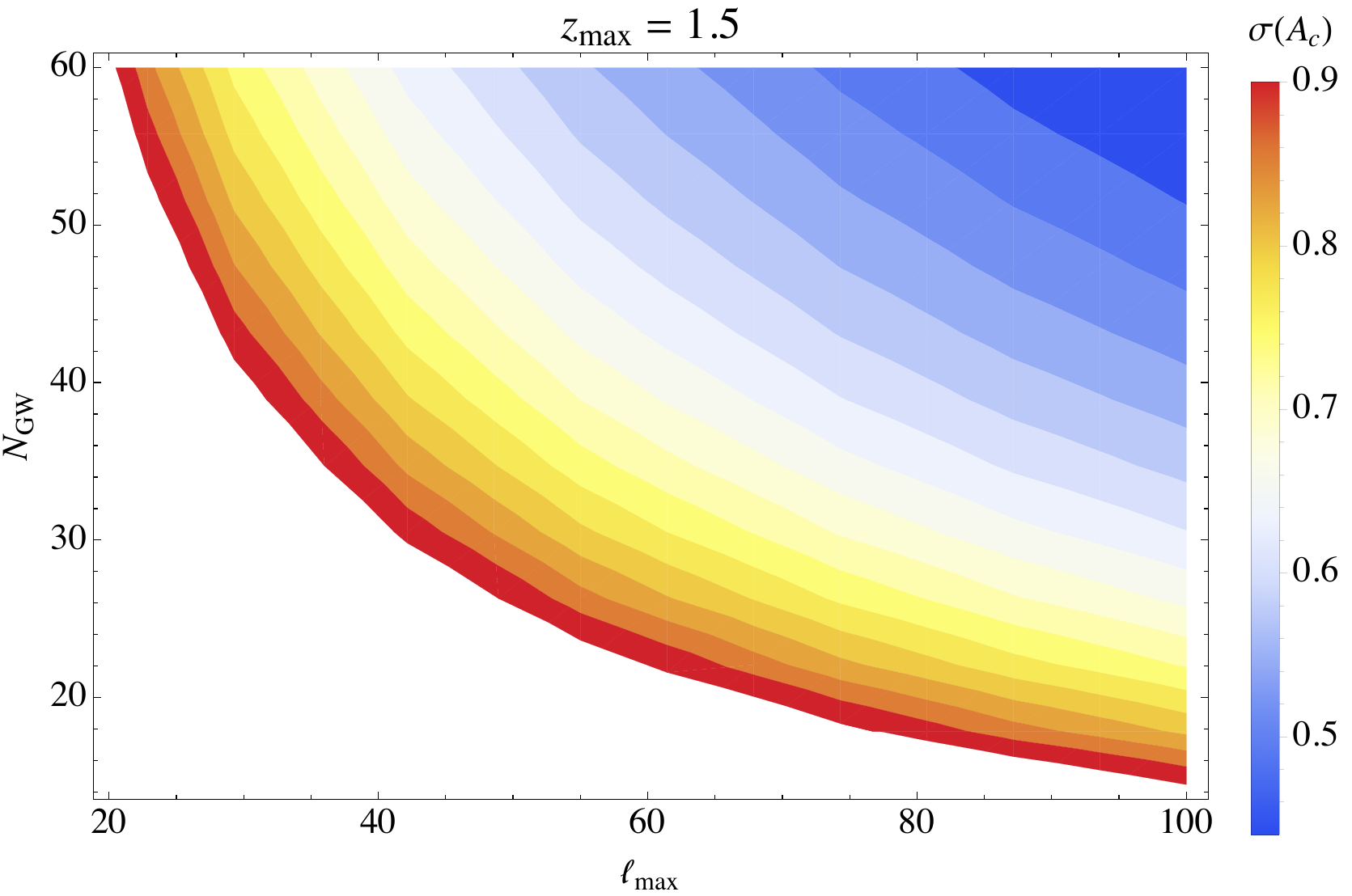}
\includegraphics[width=0.49\textwidth]{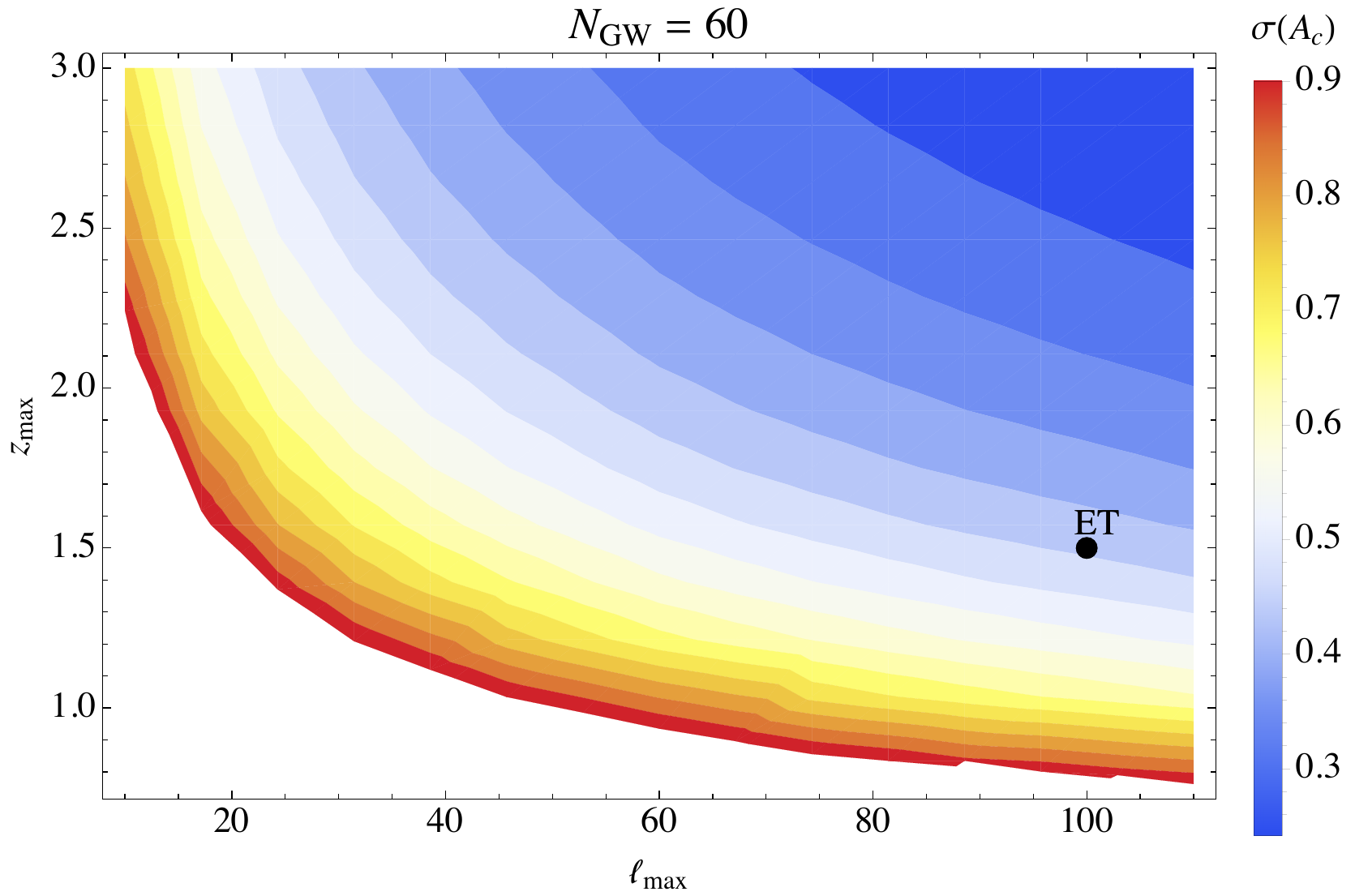}
\caption{Forecast errors on the cross-correlation amplitude $A_c$ (colorbar) as a function of different combinations 
of our free parameters.
{\it Upper Panels:} cross-correlation errors for the endpoint of stellar evolution case; {\it Lower Panels:} cross-correlation errors for the PBHs case.
The horizontal axis shows $\ell_{\rm max}$, the maximum multipole accessible. 
The left hand panels fix $z_\mathrm{max}$ and show $N_\mathrm{GW}$ on the vertical axis, while the right hand panels fix the number of events and show constraints when varying the maximum redshift observable.}  
\label{fig:dep_gen}
\end{figure*}

We now use the formalism outlined above to compute the correlation between GWs 
and galaxy catalogs. 
We study two cases separately. First, we forecast the error on the amplitude
of the cross-correlation of all GW events detected, assuming they form as stellar binaries
within galaxies. We then forecast how well one can test whether PBHs are the progenitors 
of high-mass ($\sim\!30M_\odot$) BH-binaries, by cross-correlating galaxies with only the 
higher-mass GW events.

\subsection{GW-galaxy correlation}

In the top panels of Figure~\ref{fig:dep_gen} we show the predicted error on the correlation amplitude, $\sigma_{A_c}$,
as a function of a number of parameters describing the GW
instrument used. We show results as a function of the minimum scale probed 
$\ell_{\rm max}$, the maximum redshift $z_{\rm max}$ and the number 
of BH-BH mergers detected, defined as $N_{GW} = T_{\rm obs} \mathcal{R} V_{\rm obs}$, where 
$\mathcal{R}$ is the integrated average merger rate in units of Gpc$^{-3}$ yr$^{-1}$, 
and $T_{\rm obs}$ and $V_{\rm obs}$ are the relevant observation time and volume.

It can be seen that, as expected, the main limiting factors for the detection of 
a deviation from GW-galaxy correlation are the number of GW events and the 
minimum angular scale used. In order to reduce the GW shot noise it is 
important to observe a larger volume and for a sufficient amount of time. 
For the angular scale, having more detectors will allow a better spatial 
localization and hence a larger $\ell_{\rm max}$ to be used~\cite{LIGO-net}.

\begin{figure}
\includegraphics[width=0.49\textwidth]{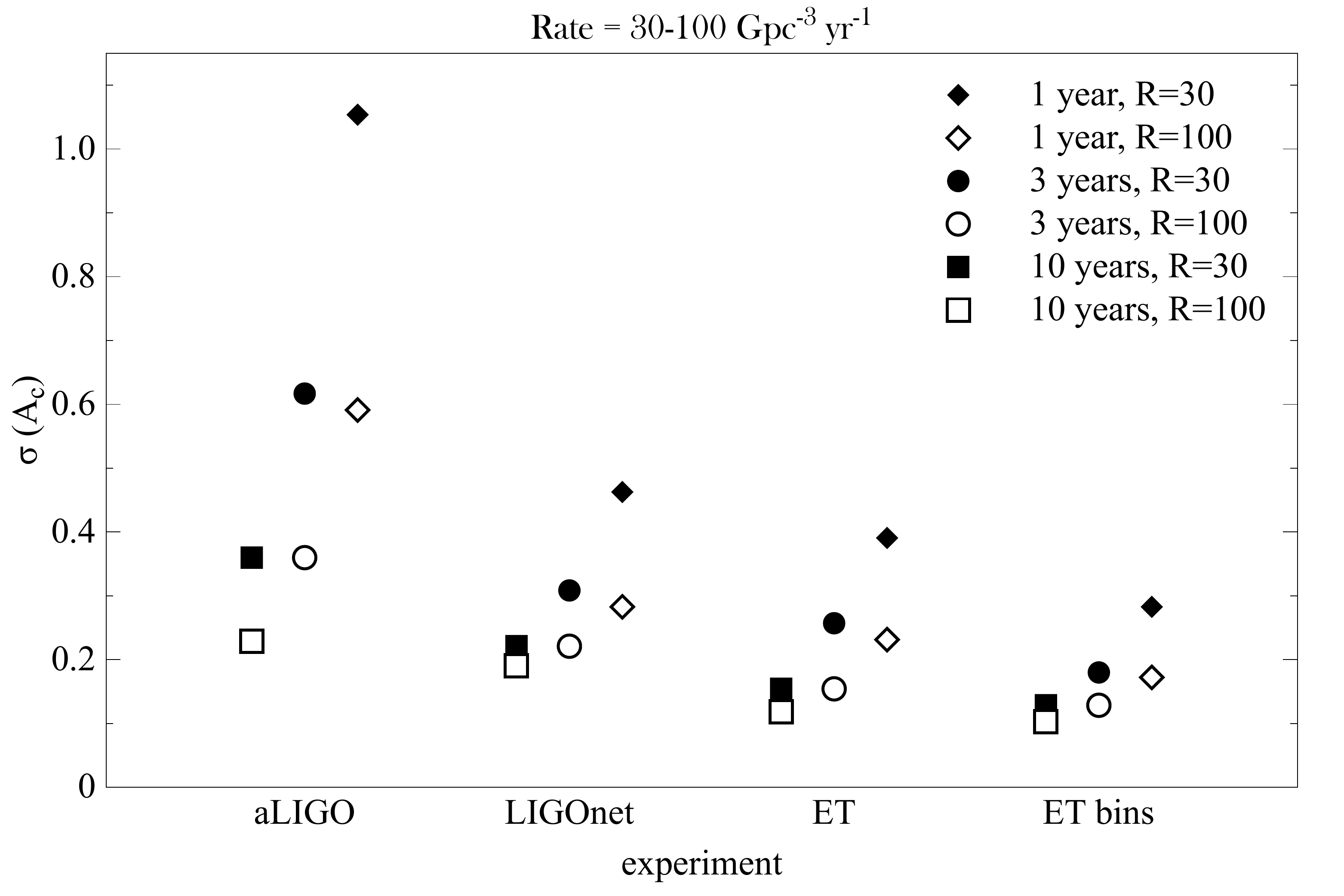}
\caption{Forecast errors on the cross-correlation amplitude, $A_c$, for different 
fiducial experiment sets, varying merger rates and years of observations. 
We show the error in the cross-correlation for all observed BH-BH mergers.
We assume the fiducial model for BHs having a stellar binary origin.
Filled symbols show the lower bound of this rate and open symbols show the upper bound.
Each column corresponds to a GW detector experiment.
}
\label{fig:exp_bGW_generic}
\end{figure}

In Figure~\ref{fig:exp_bGW_generic} we show forecasts for 
various ongoing and next-generation experiments: aLIGO (advanced LIGO), 
an extended aLIGO network (that we call LIGO-net~\cite{LIGO-net})
and the planned Einstein Telescope~\cite{Sathyaprakash:2011}, 
computing the results for the case of a single redshift bin as well as 
multiple bins, as described above. We consider observation times of 
1, 3 and 10 years. Symbols mark results for the upper and lower bounds on the 
merger rate of $30$ and $100$ Gpc$^{-3}$ yr$^{-1}$ averaged up to $z \leq 0.5$, 
and then extrapolated to higher redshifts based on the redshift-dependent $R(z)$ 
for $Z = 0.25 Z_{\odot}$ adopted from~\cite{Dominik:2013}. Note that in case 
a significant fraction of the observed GWs actually originate in high metallicity 
environments, $Z\sim Z_{\odot}$, this would mean that the GW rate will be suppressed 
at higher redshifts and ET will then not observe many more events. 

Our results indicate that instrument configurations already available 
may be able to see a hint of deviations $\Delta A_c\approx0.3$ from our fiducial 
model for galaxy progenitors, at 1-$\sigma$. Future measurements 
using the ET will yield extremely precise constraints, potentially allowing 
alternative models to be discriminated at high significance.


\subsection{Detecting PBH progenitors}

\begin{figure}
\centering
\includegraphics[width=0.49\textwidth]{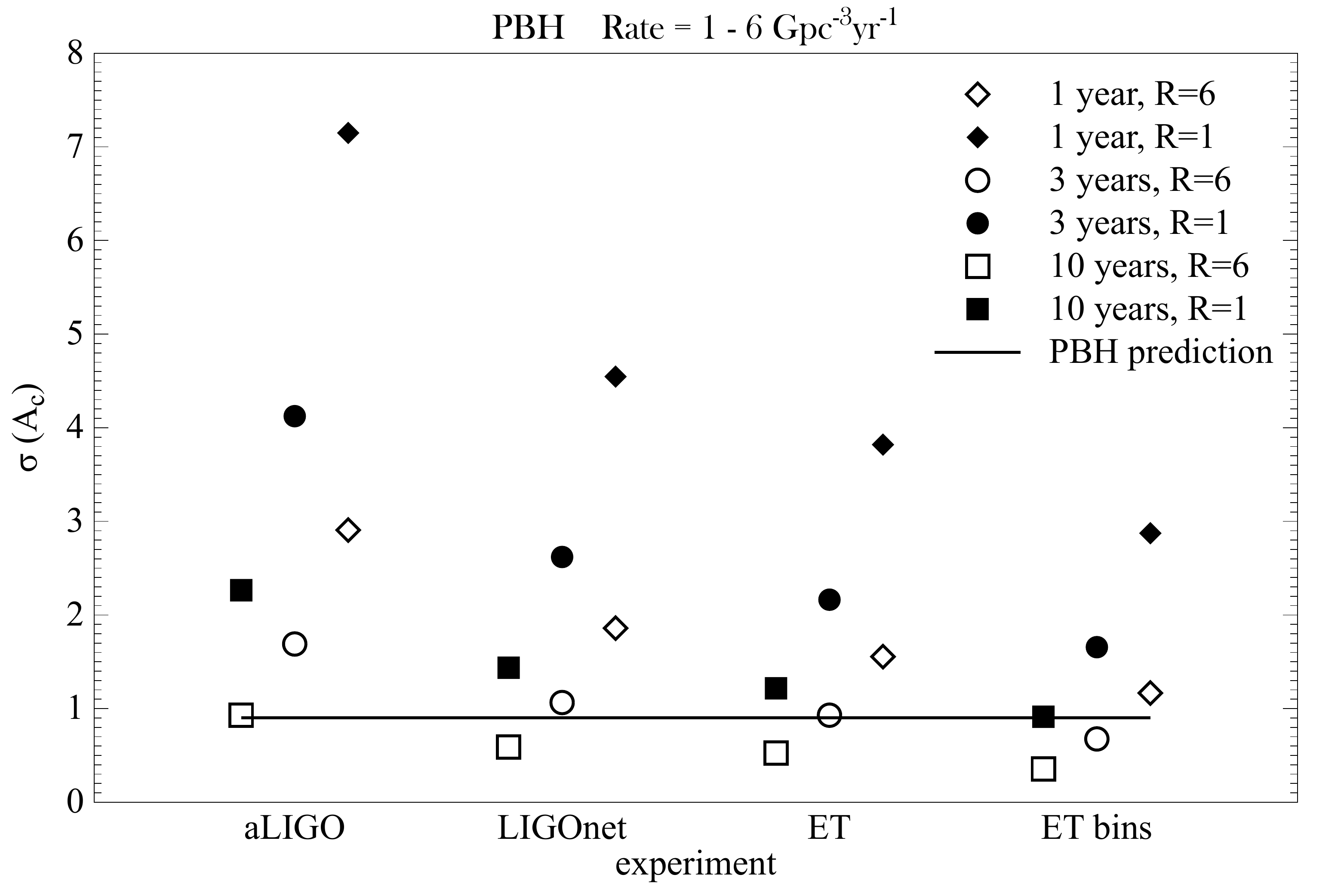}
\caption{Forecast errors on the cross-correlation amplitude, $A_c$, for different fiducial experiment sets, varying merger rates and years of observations. 
We assume the fiducial model for merger events with originating from PBHs. Filled symbols show the lower bound of this rate and open symbols show the upper bound.
Each column corresponds to a GW detector experiment.
The horizontal lines show the expected difference in the cross-correlation between a PBH and stellar binary 
origin for the BH mergers.
}
\label{fig:exp_bGW}
\end{figure}

The dependence of $\sigma(A_c)$ is shown in the lower panels of Figure~\ref{fig:dep_gen} 
for fiducial event rates and biases typical of a PBH origin for the high-mass GW events. 
The results are similar to those where the mergers originate in stellar binaries, as expected. 
The main difference is the reduced number of events, due to the fact that we use only the 
predicted high-mass BH mergers, which makes it more important to reach a better angular 
resolution and survey volume. 

In Figure~\ref{fig:exp_bGW} we show how well we can constrain a PBH origin for DM with different experiments. 
The predicted measurement precision for this model has a target threshold to cross, i.e. $\sigma(A_c) 
\lesssim b_{GW}^{\rm Stellar} - b_{GW}^{\rm PBH} = 0.9$, corresponding to the predicted difference in the correlation between 
GWs and galaxies in the PBH and stellar models. Figure~\ref{fig:exp_bGW} shows this threshold with a solid line.
Note that BH-BH mergers from stellar binaries are expected to be detectable from a wide range of BH masses, 
between 5-30 $M_\odot$~\cite{Chatterjee:2016}. Therefore, even if the dark matter is made of PBHs, a GW-event map containing only 
$M > 30 M_{\odot}$ mergers will include contributions from both primordial and
stellar BH-binaries. In this case the detection threshold would reduce accordingly to the weighted average of the difference between the biases.

We can see that if the merger rate for PBHs is at the upper end of the range considered, a 1-$\sigma$
measurement of a GW bias deviating from that of the galaxies is possible with $\approx$10 years of observations with aLIGO.
With the same merger rate, a future LIGO network could achieve the same accuracy in $\approx $3 years, 
or detect a difference in the biases at 2-$\sigma$ in 10 years. ET would increase the significance a little more. When binning the ET data, slightly more than 1 year of observation could grant a 1-$\sigma$ measurement, or 10 years could allow such detection even in the most pessimistic case for the merger rate value.
This instrument configuration would allow, for the optimistic merger rate case, a $\sim$3-$\sigma$ detection.

Comparing Figure~\ref{fig:exp_bGW_generic} to Figure \ref{fig:exp_bGW}, corresponding to detection possibilities
for PBH and stellar binary BHs, shows that the $\sigma(A_{c})$ achieved for a given GWs detector configuration is much smaller under our model for BHs originating from stellar binaries. That is mostly due to the higher assumed overall BH-BH merger rate, which leads to a smaller noise term in Eq.~\ref{eq:err-clgt}. Although the fiducial bias also changes, this makes a small difference.


\subsection{Dependence on the Galaxy Catalog}

\begin{figure}
\centering
\includegraphics[width=\columnwidth]{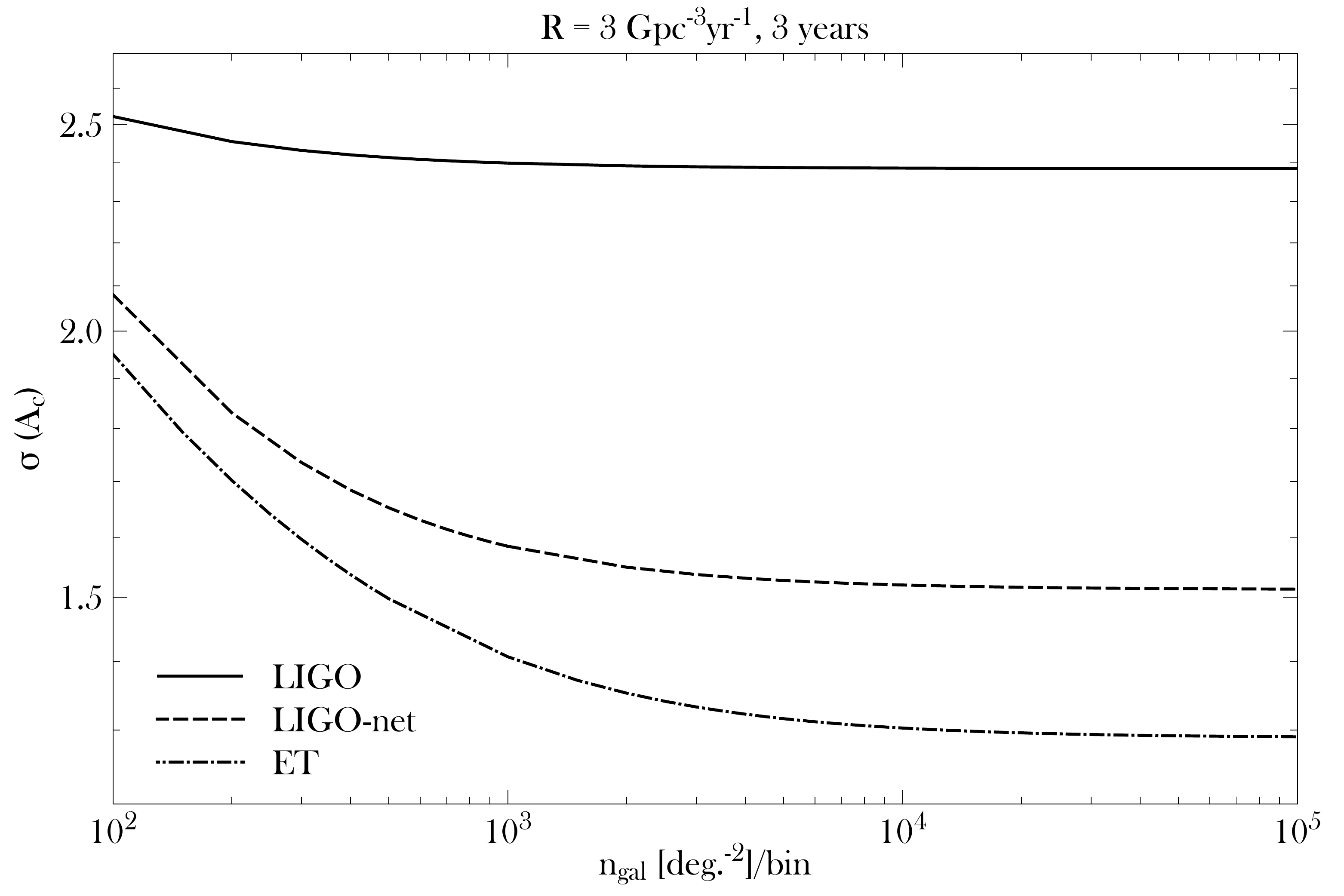}
\caption{Forecast error on the cross-correlation amplitude $A_c$, as a function of the number density of galaxies in the galaxy survey.
We show lines for three different GW detectors, each described in Section~\ref{sec:GWexp}. We assume a GW observing time of 3 years, assuming a fiducial detection rate of 3 Gpc$^{-3}$ yr$^{-1}$ for GW BH mergers.
}
\label{fig:sigma_ngal}
\end{figure}

\begin{figure}
\centering
\includegraphics[width=\columnwidth]{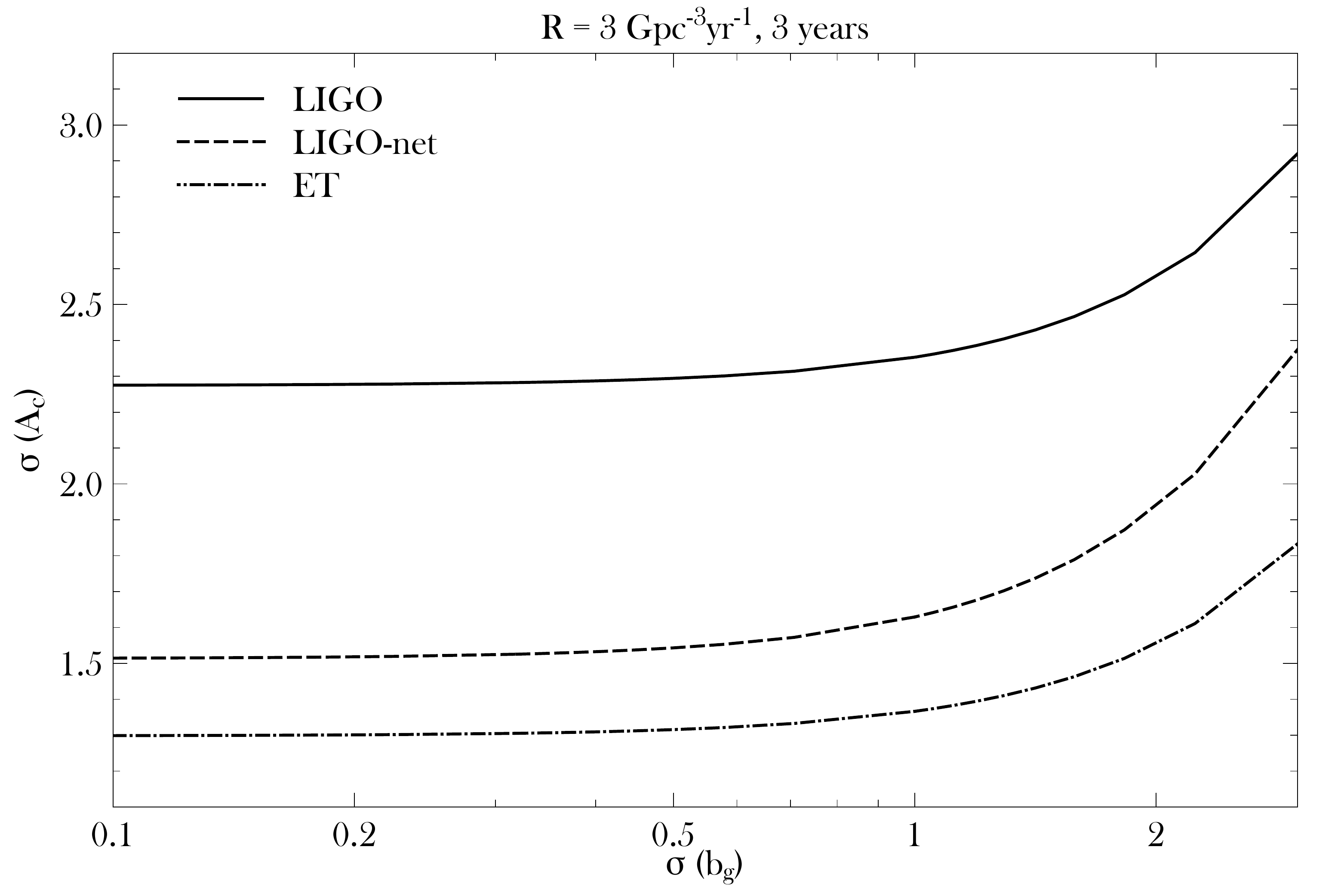}
\caption{Forecast error on on the cross-correlation amplitude $A_c$ as a function of the error in the galaxy bias (measured independently). We assume PBH progenitors with 3 years of observations for each experiment.
}
\label{fig:sigma_b_p_pbh}
\end{figure}

In Figure~\ref{fig:sigma_ngal} we show constraints on the correlation amplitude $A_c$
as a function of the number of objects in the galaxy catalog. 
It can be seen that once the number of objects reaches $\approx 
1000$ deg.$^{-2}$, the shot noise of galaxies becomes unimportant.
Of course, the range where galaxy shot noise becomes negligible 
depends on the number of GW events detected (for a small number 
of GWs, their shot noise prevents any gain by adding galaxies).
For the angular cross-correlations we are interested in, 
the best results are obtained by optimizing for number density 
and redshift range, while it is not required to obtain a precise 
redshift estimation, given that we are using projected angular correlations.
Thus photometric or radio surveys will indeed be the most useful.

We also investigate how our results will vary if we bin in redshift. 
We assume that the redshift distribution of GWs observed by the 
ET can be separated into two $z$-bins (in practice finer 
redshift-binning may be possible). We note that it is possible that specific 
models for the GW progenitors will call for particular optimal binning-strategies.

Above, we assumed that the galaxy bias will be measured to $\sim10\%$ precision 
by using the galaxy auto-correlation function. 
In Figure~\ref{fig:sigma_b_p_pbh} we investigate how the constraints depend on the error on the galaxy bias. 
We plot the error forecast as a function of the precision of measurements of 
the galaxy bias for 3 years of observation and assuming a merger rate of 3 Gpc$^{-3}$ yr$^{-1}$. 
We show that even using the lower merger rate expected for massive BH mergers, a precision of a 
few tens of percent ($\sigma(b_g)\lesssim 0.5$) will be sufficient to extract the full information 
contained in the galaxy-GW cross-correlation. Uncertainty in the galaxy bias is thus unlikely to be 
a limiting factor in practice.

\section{Discussion and Conclusions}
\label{sec:conclusions}
In this paper we have suggested that the cross-correlation of galaxy catalogs with maps
of GW-event locations can be used to statistically infer the nature of the progenitors of BH-BH 
mergers detected by current and future gravitational wave detectors. We have shown that by 
measuring the degree of cross-correlation between galaxies and 
gravitational waves, future GW experiments can potentially distinguish between GWs originating 
within galaxies and models where the merging binary systems reside 
preferentially in smaller or larger objects.

We have made forecasts for 
measurements with aLIGO in present and future configurations and with the 
planned Einstein Telescope, demonstrating under which conditions this technique may be effective.
As an example of our methodology, we presented a forecast on the 
possibility to test the hypothesis that high-mass BH-BH mergers such as GW150914 
come from the merging of PBHs of $\sim 30 M_\odot$ that could make up the dark matter 
in the Universe~\cite{Bird:2016}. 
Since in this model the vast majority of mergers occur in low-mass halos, the sources 
of GW events should be more uniformly distributed on the sky, with a low bias, and 
with an almost flat redshift distribution.
Our results show that aLIGO  $+$ VIRGO may be able to probe this model after $\approx$10 years of observations,
under optimistic assumptions on the resulting GW event rate.
A future LIGO network including new detectors would 
be able to test this model with an increased precision (over a similar observing time), while the ET should allow a 
measurement at marginal significance even in the case of a low merger rate and a 
relatively poor determination of the bias of galaxies.

We emphasize that our predictions were derived under fairly conservative 
assumptions; as noted above, having a galaxy bias that increases with redshift 
would make it easier to detect the PBHs scenario, by increasing $\Delta b$. For 
the ET case, we assumed a conservative minimum angular scale of $100$ and 
even more conservatively, a maximum redshift $z_{\rm max}=1.5$. Clearly, increasing 
the maximum $z$ probed will increase the number of events observed, hence 
increasing the constraining power of the instrument. Finally, much better results 
could be obtained with proposed future instruments such as DECIGO~\cite{decigo}.

For the cases considered above, we have shown that the properties of 
the galaxy survey used is not a limiting factor.  
More generally, we note that specific models for the progenitors of BH-BH 
mergers can in principle predict deviations from the standard
case of stellar progenitors in several parts of the parameter space, 
i.e. bias, redshift range and angular scales.
To probe these models, a variety of galaxy surveys that are planned for the next few years will be available, so one 
could choose to use a narrow and deep observation (by using e.g. PFS~\cite{pfs}) 
or a shallower but full-sky one (e.g. SPHEREx~\cite{spherex}).

Finally, our methodology is focused on determining the nature of binary 
progenitors by making use of the cross-correlation of galaxy number counts 
with GW events. It is worth noting that auto and cross-correlations of GW maps can also be used in principle 
to constrain cosmological parameters, using observables such as weak gravitational lensing~\cite{Cutler:2009, Camera:2013} 
and the cross-correlation of GW maps with CMB temperature maps might enable to detect the Integrated Sachs-Wolfe effect~\cite{Laguna:2009} (further cosmological investigations have been recently proposed in~\cite{Oguri:2016, Collett:2016}).

\vspace{0.3cm}

{\bf Acknowledgments} \\
The authors thank Yacine Ali-Ha\"imoud, Daniele Bertacca, Vincent Desjacques, Juan Garcia-Bellido, Marc Kamionkowski, Cristiano Porciani, Sabino Matarrese, and Eleonora Villa for useful discussions. This work was supported by NSF Grant No. 0244990, NASA NNX15AB18G, the 
John Templeton Foundation, and the Simons Foundation.
SB was supported by NASA through Einstein Postdoctoral Fellowship Award Number PF5-160133. 

\vspace{-0.2in}

\bibliography{CCF-GW_sub}

\begin{thebibliography}{58}%
\makeatletter
\providecommand \@ifxundefined [1]{%
 \@ifx{#1\undefined}
}%
\providecommand \@ifnum [1]{%
 \ifnum #1\expandafter \@firstoftwo
 \else \expandafter \@secondoftwo
 \fi
}%
\providecommand \@ifx [1]{%
 \ifx #1\expandafter \@firstoftwo
 \else \expandafter \@secondoftwo
 \fi
}%
\providecommand \natexlab [1]{#1}%
\providecommand \enquote  [1]{``#1''}%
\providecommand \bibnamefont  [1]{#1}%
\providecommand \bibfnamefont [1]{#1}%
\providecommand \citenamefont [1]{#1}%
\providecommand \href@noop [0]{\@secondoftwo}%
\providecommand \href [0]{\begingroup \@sanitize@url \@href}%
\providecommand \@href[1]{\@@startlink{#1}\@@href}%
\providecommand \@@href[1]{\endgroup#1\@@endlink}%
\providecommand \@sanitize@url [0]{\catcode `\\12\catcode `\$12\catcode
  `\&12\catcode `\#12\catcode `\^12\catcode `\_12\catcode `\%12\relax}%
\providecommand \@@startlink[1]{}%
\providecommand \@@endlink[0]{}%
\providecommand \url  [0]{\begingroup\@sanitize@url \@url }%
\providecommand \@url [1]{\endgroup\@href {#1}{\urlprefix }}%
\providecommand \urlprefix  [0]{URL }%
\providecommand \Eprint [0]{\href }%
\providecommand \doibase [0]{http://dx.doi.org/}%
\providecommand \selectlanguage [0]{\@gobble}%
\providecommand \bibinfo  [0]{\@secondoftwo}%
\providecommand \bibfield  [0]{\@secondoftwo}%
\providecommand \translation [1]{[#1]}%
\providecommand \BibitemOpen [0]{}%
\providecommand \bibitemStop [0]{}%
\providecommand \bibitemNoStop [0]{.\EOS\space}%
\providecommand \EOS [0]{\spacefactor3000\relax}%
\providecommand \BibitemShut  [1]{\csname bibitem#1\endcsname}%
\let\auto@bib@innerbib\@empty
\bibitem [{\citenamefont {Collaboration}\ and\ \citenamefont
  {Collaboration}(2016)}]{LIGO:GW}%
  \BibitemOpen
  \bibfield  {author} {\bibinfo {author} {\bibfnamefont {L.~S.}\ \bibnamefont
  {Collaboration}}\ and\ \bibinfo {author} {\bibfnamefont {V.}~\bibnamefont
  {Collaboration}},\ }\href {http://arxiv.org/abs/1602.03837} {\bibfield
  {journal} {\bibinfo  {journal} {Phys. Rev. Lett.}\ }\textbf {\bibinfo
  {volume} {116}},\ \bibinfo {pages} {061102} (\bibinfo {year} {2016})},\
  \Eprint {http://arxiv.org/abs/1602.03837} {1602.03837} \BibitemShut {NoStop}%
\bibitem [{\citenamefont {Connaughton}\ \emph {et~al.}(2016)\citenamefont
  {Connaughton} \emph {et~al.}}]{Connaughton:2016umz}%
  \BibitemOpen
  \bibfield  {author} {\bibinfo {author} {\bibfnamefont {V.}~\bibnamefont
  {Connaughton}} \emph {et~al.},\ }\href@noop {} {\  (\bibinfo {year}
  {2016})},\ \Eprint {http://arxiv.org/abs/1602.03920} {arXiv:1602.03920
  [astro-ph.HE]} \BibitemShut {NoStop}%
\bibitem [{\citenamefont {Loeb}(2016)}]{Loeb:2016fzn}%
  \BibitemOpen
  \bibfield  {author} {\bibinfo {author} {\bibfnamefont {A.}~\bibnamefont
  {Loeb}},\ }\href {\doibase 10.3847/2041-8205/819/2/L21} {\bibfield  {journal}
  {\bibinfo  {journal} {Astrophys. J.}\ }\textbf {\bibinfo {volume} {819}},\
  \bibinfo {pages} {L21} (\bibinfo {year} {2016})},\ \Eprint
  {http://arxiv.org/abs/1602.04735} {arXiv:1602.04735 [astro-ph.HE]}
  \BibitemShut {NoStop}%
\bibitem [{\citenamefont {Kotera}\ and\ \citenamefont
  {Silk}(2016)}]{Kotera:2016dmp}%
  \BibitemOpen
  \bibfield  {author} {\bibinfo {author} {\bibfnamefont {K.}~\bibnamefont
  {Kotera}}\ and\ \bibinfo {author} {\bibfnamefont {J.}~\bibnamefont {Silk}},\
  }\href@noop {} {\  (\bibinfo {year} {2016})},\ \Eprint
  {http://arxiv.org/abs/1602.06961} {arXiv:1602.06961 [astro-ph.HE]}
  \BibitemShut {NoStop}%
\bibitem [{\citenamefont {Chatterjee}\ \emph {et~al.}(2016)\citenamefont
  {Chatterjee}, \citenamefont {Rodriguez},\ and\ \citenamefont
  {Rasio}}]{Chatterjee:2016}%
  \BibitemOpen
  \bibfield  {author} {\bibinfo {author} {\bibfnamefont {S.}~\bibnamefont
  {Chatterjee}}, \bibinfo {author} {\bibfnamefont {C.~L.}\ \bibnamefont
  {Rodriguez}}, \ and\ \bibinfo {author} {\bibfnamefont {F.~A.}\ \bibnamefont
  {Rasio}},\ }\href {http://arxiv.org/abs/1603.00884} {\  (\bibinfo {year}
  {2016})},\ \Eprint {http://arxiv.org/abs/1603.00884} {1603.00884}
  \BibitemShut {NoStop}%
\bibitem [{\citenamefont {Namikawa}\ \emph
  {et~al.}(2016{\natexlab{a}})\citenamefont {Namikawa}, \citenamefont
  {Nishizawa},\ and\ \citenamefont {Taruya}}]{Namikawa:2016}%
  \BibitemOpen
  \bibfield  {author} {\bibinfo {author} {\bibfnamefont {T.}~\bibnamefont
  {Namikawa}}, \bibinfo {author} {\bibfnamefont {A.}~\bibnamefont {Nishizawa}},
  \ and\ \bibinfo {author} {\bibfnamefont {A.}~\bibnamefont {Taruya}},\ }\href
  {http://arxiv.org/abs/1603.08072} {\  (\bibinfo {year}
  {2016}{\natexlab{a}})},\ \Eprint {http://arxiv.org/abs/1603.08072}
  {1603.08072} \BibitemShut {NoStop}%
\bibitem [{\citenamefont {Bird}\ \emph {et~al.}(2016)\citenamefont {Bird},
  \citenamefont {Cholis}, \citenamefont {Mu{\~n}oz}, \citenamefont
  {Ali-Ha{\"\i}moud}, \citenamefont {Kamionkowski}, \citenamefont {Kovetz},
  \citenamefont {Raccanelli},\ and\ \citenamefont {Riess}}]{Bird:2016}%
  \BibitemOpen
  \bibfield  {author} {\bibinfo {author} {\bibfnamefont {S.}~\bibnamefont
  {Bird}}, \bibinfo {author} {\bibfnamefont {I.}~\bibnamefont {Cholis}},
  \bibinfo {author} {\bibfnamefont {J.~B.}\ \bibnamefont {Mu{\~n}oz}}, \bibinfo
  {author} {\bibfnamefont {Y.}~\bibnamefont {Ali-Ha{\"\i}moud}}, \bibinfo
  {author} {\bibfnamefont {M.}~\bibnamefont {Kamionkowski}}, \bibinfo {author}
  {\bibfnamefont {E.~D.}\ \bibnamefont {Kovetz}}, \bibinfo {author}
  {\bibfnamefont {A.}~\bibnamefont {Raccanelli}}, \ and\ \bibinfo {author}
  {\bibfnamefont {A.~G.}\ \bibnamefont {Riess}},\ }\href
  {http://arxiv.org/abs/1603.00464} {\  (\bibinfo {year} {2016})},\ \Eprint
  {http://arxiv.org/abs/1603.00464} {1603.00464} \BibitemShut {NoStop}%
\bibitem [{\citenamefont {Garcia-Bellido}\ \emph {et~al.}(1996)\citenamefont
  {Garcia-Bellido}, \citenamefont {Linde},\ and\ \citenamefont
  {Wands}}]{Garcia-Bellido:1996}%
  \BibitemOpen
  \bibfield  {author} {\bibinfo {author} {\bibfnamefont {J.}~\bibnamefont
  {Garcia-Bellido}}, \bibinfo {author} {\bibfnamefont {A.}~\bibnamefont
  {Linde}}, \ and\ \bibinfo {author} {\bibfnamefont {D.}~\bibnamefont
  {Wands}},\ }\href {http://arxiv.org/abs/astro-ph/9605094} {\bibfield
  {journal} {\bibinfo  {journal} {Phys.Rev.D}\ }\textbf {\bibinfo {volume}
  {54}},\ \bibinfo {pages} {6040} (\bibinfo {year} {1996})},\ \Eprint
  {http://arxiv.org/abs/astro-ph/9605094} {astro-ph/9605094} \BibitemShut
  {NoStop}%
\bibitem [{\citenamefont {Nakamura}\ \emph {et~al.}(1997)\citenamefont
  {Nakamura}, \citenamefont {Sasaki}, \citenamefont {Tanaka},\ and\
  \citenamefont {Thorne}}]{Nakamura:1997}%
  \BibitemOpen
  \bibfield  {author} {\bibinfo {author} {\bibfnamefont {T.}~\bibnamefont
  {Nakamura}}, \bibinfo {author} {\bibfnamefont {M.}~\bibnamefont {Sasaki}},
  \bibinfo {author} {\bibfnamefont {T.}~\bibnamefont {Tanaka}}, \ and\ \bibinfo
  {author} {\bibfnamefont {K.~S.}\ \bibnamefont {Thorne}},\ }\href
  {http://arxiv.org/abs/astro-ph/9708060} {\bibfield  {journal} {\bibinfo
  {journal} {Astrophys.J.}\ }\textbf {\bibinfo {volume} {487}},\ \bibinfo
  {pages} {L139} (\bibinfo {year} {1997})},\ \Eprint
  {http://arxiv.org/abs/astro-ph/9708060} {astro-ph/9708060} \BibitemShut
  {NoStop}%
\bibitem [{\citenamefont {Clesse}\ and\ \citenamefont
  {Garc{\'\i}a-Bellido}(2016)}]{Clesse:2016}%
  \BibitemOpen
  \bibfield  {author} {\bibinfo {author} {\bibfnamefont {S.}~\bibnamefont
  {Clesse}}\ and\ \bibinfo {author} {\bibfnamefont {J.}~\bibnamefont
  {Garc{\'\i}a-Bellido}},\ }\href {http://arxiv.org/abs/1603.05234} {\
  (\bibinfo {year} {2016})},\ \Eprint {http://arxiv.org/abs/1603.05234}
  {1603.05234} \BibitemShut {NoStop}%
\bibitem [{\citenamefont {Sasaki}\ \emph {et~al.}(2016)\citenamefont {Sasaki},
  \citenamefont {Suyama}, \citenamefont {Tanaka},\ and\ \citenamefont
  {Yokoyama}}]{Sasaki:2016}%
  \BibitemOpen
  \bibfield  {author} {\bibinfo {author} {\bibfnamefont {M.}~\bibnamefont
  {Sasaki}}, \bibinfo {author} {\bibfnamefont {T.}~\bibnamefont {Suyama}},
  \bibinfo {author} {\bibfnamefont {T.}~\bibnamefont {Tanaka}}, \ and\ \bibinfo
  {author} {\bibfnamefont {S.}~\bibnamefont {Yokoyama}},\ }\href
  {http://arxiv.org/abs/1603.08338} {\  (\bibinfo {year} {2016})},\ \Eprint
  {http://arxiv.org/abs/1603.08338} {1603.08338} \BibitemShut {NoStop}%
\bibitem [{\citenamefont {Kinugawa}\ \emph {et~al.}(2014)\citenamefont
  {Kinugawa}, \citenamefont {Inayoshi}, \citenamefont {Hotokezaka},
  \citenamefont {Nakauchi},\ and\ \citenamefont {Nakamura}}]{Kinugawa:2014zha}%
  \BibitemOpen
  \bibfield  {author} {\bibinfo {author} {\bibfnamefont {T.}~\bibnamefont
  {Kinugawa}}, \bibinfo {author} {\bibfnamefont {K.}~\bibnamefont {Inayoshi}},
  \bibinfo {author} {\bibfnamefont {K.}~\bibnamefont {Hotokezaka}}, \bibinfo
  {author} {\bibfnamefont {D.}~\bibnamefont {Nakauchi}}, \ and\ \bibinfo
  {author} {\bibfnamefont {T.}~\bibnamefont {Nakamura}},\ }\href {\doibase
  10.1093/mnras/stu1022} {\bibfield  {journal} {\bibinfo  {journal} {Mon. Not.
  Roy. Astron. Soc.}\ }\textbf {\bibinfo {volume} {442}},\ \bibinfo {pages}
  {2963} (\bibinfo {year} {2014})},\ \Eprint {http://arxiv.org/abs/1402.6672}
  {arXiv:1402.6672 [astro-ph.HE]} \BibitemShut {NoStop}%
\bibitem [{\citenamefont {Inayoshi}\ \emph {et~al.}(2016)\citenamefont
  {Inayoshi}, \citenamefont {Kashiyama}, \citenamefont {Visbal},\ and\
  \citenamefont {Haiman}}]{Inayoshi:2016hco}%
  \BibitemOpen
  \bibfield  {author} {\bibinfo {author} {\bibfnamefont {K.}~\bibnamefont
  {Inayoshi}}, \bibinfo {author} {\bibfnamefont {K.}~\bibnamefont {Kashiyama}},
  \bibinfo {author} {\bibfnamefont {E.}~\bibnamefont {Visbal}}, \ and\ \bibinfo
  {author} {\bibfnamefont {Z.}~\bibnamefont {Haiman}},\ }\href@noop {} {\
  (\bibinfo {year} {2016})},\ \Eprint {http://arxiv.org/abs/1603.06921}
  {arXiv:1603.06921 [astro-ph.GA]} \BibitemShut {NoStop}%
\bibitem [{\citenamefont {Hartwig}\ \emph {et~al.}(2016)\citenamefont
  {Hartwig}, \citenamefont {Volonteri}, \citenamefont {Bromm}, \citenamefont
  {Klessen}, \citenamefont {Barausse}, \citenamefont {Magg},\ and\
  \citenamefont {Stacy}}]{Hartwig:2016nde}%
  \BibitemOpen
  \bibfield  {author} {\bibinfo {author} {\bibfnamefont {T.}~\bibnamefont
  {Hartwig}}, \bibinfo {author} {\bibfnamefont {M.}~\bibnamefont {Volonteri}},
  \bibinfo {author} {\bibfnamefont {V.}~\bibnamefont {Bromm}}, \bibinfo
  {author} {\bibfnamefont {R.~S.}\ \bibnamefont {Klessen}}, \bibinfo {author}
  {\bibfnamefont {E.}~\bibnamefont {Barausse}}, \bibinfo {author}
  {\bibfnamefont {M.}~\bibnamefont {Magg}}, \ and\ \bibinfo {author}
  {\bibfnamefont {A.}~\bibnamefont {Stacy}},\ }\href@noop {} {\  (\bibinfo
  {year} {2016})},\ \Eprint {http://arxiv.org/abs/1603.05655} {arXiv:1603.05655
  [astro-ph.GA]} \BibitemShut {NoStop}%
\bibitem [{\citenamefont {O'Leary}\ \emph {et~al.}(2008)\citenamefont
  {O'Leary}, \citenamefont {Kocsis},\ and\ \citenamefont {Loeb}}]{OLeary:2008}%
  \BibitemOpen
  \bibfield  {author} {\bibinfo {author} {\bibfnamefont {R.~M.}\ \bibnamefont
  {O'Leary}}, \bibinfo {author} {\bibfnamefont {B.}~\bibnamefont {Kocsis}}, \
  and\ \bibinfo {author} {\bibfnamefont {A.}~\bibnamefont {Loeb}},\ }\href
  {http://arxiv.org/abs/0807.2638} {\  (\bibinfo {year} {2008})},\ \Eprint
  {http://arxiv.org/abs/0807.2638} {0807.2638} \BibitemShut {NoStop}%
\bibitem [{\citenamefont {Stone}\ \emph {et~al.}(2013)\citenamefont {Stone},
  \citenamefont {Sari},\ and\ \citenamefont {Loeb}}]{Stone:2012uk}%
  \BibitemOpen
  \bibfield  {author} {\bibinfo {author} {\bibfnamefont {N.}~\bibnamefont
  {Stone}}, \bibinfo {author} {\bibfnamefont {R.}~\bibnamefont {Sari}}, \ and\
  \bibinfo {author} {\bibfnamefont {A.}~\bibnamefont {Loeb}},\ }\href {\doibase
  10.1093/mnras/stt1270} {\bibfield  {journal} {\bibinfo  {journal} {Mon. Not.
  Roy. Astron. Soc.}\ }\textbf {\bibinfo {volume} {435}},\ \bibinfo {pages}
  {1809} (\bibinfo {year} {2013})},\ \Eprint {http://arxiv.org/abs/1210.3374}
  {arXiv:1210.3374 [astro-ph.HE]} \BibitemShut {NoStop}%
\bibitem [{\citenamefont {Ali-Ha{\"\i}moud}\ \emph {et~al.}(2016)\citenamefont
  {Ali-Ha{\"\i}moud}, \citenamefont {Kovetz},\ and\ \citenamefont
  {Silk}}]{Ali-Haimoud:2015bfg}%
  \BibitemOpen
  \bibfield  {author} {\bibinfo {author} {\bibfnamefont {Y.}~\bibnamefont
  {Ali-Ha{\"\i}moud}}, \bibinfo {author} {\bibfnamefont {E.~D.}\ \bibnamefont
  {Kovetz}}, \ and\ \bibinfo {author} {\bibfnamefont {J.}~\bibnamefont
  {Silk}},\ }\href {\doibase 10.1103/PhysRevD.93.043508} {\bibfield  {journal}
  {\bibinfo  {journal} {Phys. Rev.}\ }\textbf {\bibinfo {volume} {D93}},\
  \bibinfo {pages} {043508} (\bibinfo {year} {2016})},\ \Eprint
  {http://arxiv.org/abs/1511.02232} {arXiv:1511.02232 [astro-ph.HE]}
  \BibitemShut {NoStop}%
\bibitem [{\citenamefont {B.Sathyaprakash}\ \emph {et~al.}(2011)\citenamefont
  {B.Sathyaprakash}, \citenamefont {M.Abernathy}, \citenamefont {F.Acernese},
  \citenamefont {P.Amaro-Seoane}, \citenamefont {N.Andersson}, \citenamefont
  {K.Arun}, \citenamefont {F.Barone}, \citenamefont {B.Barr}, \citenamefont
  {M.Barsuglia}, \citenamefont {M.Beker}, \citenamefont {N.Beveridge},
  \citenamefont {S.Birindelli}, \citenamefont {S.Bose}, \citenamefont {L.Bosi},
  \citenamefont {S.Braccini}, \citenamefont {C.Bradaschia}, \citenamefont
  {T.Bulik}, \citenamefont {E.Calloni}, \citenamefont {G.Cella}, \citenamefont
  {E.Chassande-Mottin}, \citenamefont {S.Chelkowski}, \citenamefont
  {A.Chincarini}, \citenamefont {J.Clark}, \citenamefont {E.Coccia},
  \citenamefont {C.Colacino}, \citenamefont {J.Colas}, \citenamefont
  {A.Cumming}, \citenamefont {L.Cunningham}, \citenamefont {E.Cuoco},
  \citenamefont {S.Danilishin}, \citenamefont {K.Danzmann}, \citenamefont
  {R.De.Salvo}, \citenamefont {T.Dent}, \citenamefont {R.De.Rosa},
  \citenamefont {L.Di.Fiore}, \citenamefont {A.Di.Virgilio}, \citenamefont
  {M.Doets}, \citenamefont {V.Fafone}, \citenamefont {P.Falferi}, \citenamefont
  {R.Flaminio}, \citenamefont {J.Franc}, \citenamefont {F.Frasconi},
  \citenamefont {A.Freise}, \citenamefont {D.Friedrich}, \citenamefont
  {P.Fulda}, \citenamefont {J.Gair}, \citenamefont {G.Gemme}, \citenamefont
  {E.Genin}, \citenamefont {A.Gennai}, \citenamefont {A.Giazotto},
  \citenamefont {K.Glampedakis}, \citenamefont {C.Gr{\"a}f}, \citenamefont
  {M.Granata}, \citenamefont {H.Grote}, \citenamefont {G.Guidi}, \citenamefont
  {A.Gurkovsky}, \citenamefont {G.Hammond}, \citenamefont {M.Hannam},
  \citenamefont {J.Harms}, \citenamefont {D.Heinert}, \citenamefont {M.Hendry},
  \citenamefont {I.Heng}, \citenamefont {E.Hennes}, \citenamefont {S.Hild},
  \citenamefont {J.Hough}, \citenamefont {S.Husa}, \citenamefont {S.Huttner},
  \citenamefont {G.Jones}, \citenamefont {F.Khalili}, \citenamefont
  {K.Kokeyama}, \citenamefont {K.Kokkotas}, \citenamefont {B.Krishnan},\ and\
  \citenamefont {T.G.F.Li}}]{Sathyaprakash:2011}%
  \BibitemOpen
  \bibfield  {author} {\bibinfo {author} {\bibnamefont {B.Sathyaprakash}},
  \bibinfo {author} {\bibnamefont {M.Abernathy}}, \bibinfo {author}
  {\bibnamefont {F.Acernese}}, \bibinfo {author} {\bibnamefont
  {P.Amaro-Seoane}}, \bibinfo {author} {\bibnamefont {N.Andersson}}, \bibinfo
  {author} {\bibnamefont {K.Arun}}, \bibinfo {author} {\bibnamefont
  {F.Barone}}, \bibinfo {author} {\bibnamefont {B.Barr}}, \bibinfo {author}
  {\bibnamefont {M.Barsuglia}}, \bibinfo {author} {\bibnamefont {M.Beker}},
  \bibinfo {author} {\bibnamefont {N.Beveridge}}, \bibinfo {author}
  {\bibnamefont {S.Birindelli}}, \bibinfo {author} {\bibnamefont {S.Bose}},
  \bibinfo {author} {\bibnamefont {L.Bosi}}, \bibinfo {author} {\bibnamefont
  {S.Braccini}}, \bibinfo {author} {\bibnamefont {C.Bradaschia}}, \bibinfo
  {author} {\bibnamefont {T.Bulik}}, \bibinfo {author} {\bibnamefont
  {E.Calloni}}, \bibinfo {author} {\bibnamefont {G.Cella}}, \bibinfo {author}
  {\bibnamefont {E.Chassande-Mottin}}, \bibinfo {author} {\bibnamefont
  {S.Chelkowski}}, \bibinfo {author} {\bibnamefont {A.Chincarini}}, \bibinfo
  {author} {\bibnamefont {J.Clark}}, \bibinfo {author} {\bibnamefont
  {E.Coccia}}, \bibinfo {author} {\bibnamefont {C.Colacino}}, \bibinfo {author}
  {\bibnamefont {J.Colas}}, \bibinfo {author} {\bibnamefont {A.Cumming}},
  \bibinfo {author} {\bibnamefont {L.Cunningham}}, \bibinfo {author}
  {\bibnamefont {E.Cuoco}}, \bibinfo {author} {\bibnamefont {S.Danilishin}},
  \bibinfo {author} {\bibnamefont {K.Danzmann}}, \bibinfo {author}
  {\bibnamefont {R.De.Salvo}}, \bibinfo {author} {\bibnamefont {T.Dent}},
  \bibinfo {author} {\bibnamefont {R.De.Rosa}}, \bibinfo {author} {\bibnamefont
  {L.Di.Fiore}}, \bibinfo {author} {\bibnamefont {A.Di.Virgilio}}, \bibinfo
  {author} {\bibnamefont {M.Doets}}, \bibinfo {author} {\bibnamefont
  {V.Fafone}}, \bibinfo {author} {\bibnamefont {P.Falferi}}, \bibinfo {author}
  {\bibnamefont {R.Flaminio}}, \bibinfo {author} {\bibnamefont {J.Franc}},
  \bibinfo {author} {\bibnamefont {F.Frasconi}}, \bibinfo {author}
  {\bibnamefont {A.Freise}}, \bibinfo {author} {\bibnamefont {D.Friedrich}},
  \bibinfo {author} {\bibnamefont {P.Fulda}}, \bibinfo {author} {\bibnamefont
  {J.Gair}}, \bibinfo {author} {\bibnamefont {G.Gemme}}, \bibinfo {author}
  {\bibnamefont {E.Genin}}, \bibinfo {author} {\bibnamefont {A.Gennai}},
  \bibinfo {author} {\bibnamefont {A.Giazotto}}, \bibinfo {author}
  {\bibnamefont {K.Glampedakis}}, \bibinfo {author} {\bibnamefont
  {C.Gr{\"a}f}}, \bibinfo {author} {\bibnamefont {M.Granata}}, \bibinfo
  {author} {\bibnamefont {H.Grote}}, \bibinfo {author} {\bibnamefont
  {G.Guidi}}, \bibinfo {author} {\bibnamefont {A.Gurkovsky}}, \bibinfo {author}
  {\bibnamefont {G.Hammond}}, \bibinfo {author} {\bibnamefont {M.Hannam}},
  \bibinfo {author} {\bibnamefont {J.Harms}}, \bibinfo {author} {\bibnamefont
  {D.Heinert}}, \bibinfo {author} {\bibnamefont {M.Hendry}}, \bibinfo {author}
  {\bibnamefont {I.Heng}}, \bibinfo {author} {\bibnamefont {E.Hennes}},
  \bibinfo {author} {\bibnamefont {S.Hild}}, \bibinfo {author} {\bibnamefont
  {J.Hough}}, \bibinfo {author} {\bibnamefont {S.Husa}}, \bibinfo {author}
  {\bibnamefont {S.Huttner}}, \bibinfo {author} {\bibnamefont {G.Jones}},
  \bibinfo {author} {\bibnamefont {F.Khalili}}, \bibinfo {author} {\bibnamefont
  {K.Kokeyama}}, \bibinfo {author} {\bibnamefont {K.Kokkotas}}, \bibinfo
  {author} {\bibnamefont {B.Krishnan}}, \ and\ \bibinfo {author} {\bibnamefont
  {T.G.F.Li}},\ }\href {http://arxiv.org/abs/1108.1423} {\bibfield  {journal}
  {\bibinfo  {journal} {2011 Gravitational Waves and Experimental Gravity, eds
  Etienne Auge and Jacques Dumarchez and Jean Tran Thanh Van, The Gioi
  Publishers, Vietnam}\ } (\bibinfo {year} {2011})},\ \Eprint
  {http://arxiv.org/abs/1108.1423} {1108.1423} \BibitemShut {NoStop}%
\bibitem [{\citenamefont {Raccanelli}\ \emph {et~al.}(2008)\citenamefont
  {Raccanelli}, \citenamefont {Bonaldi}, \citenamefont {Negrello},
  \citenamefont {Matarrese}, \citenamefont {Tormen},\ and\ \citenamefont
  {Zotti}}]{Raccanelli:2008}%
  \BibitemOpen
  \bibfield  {author} {\bibinfo {author} {\bibfnamefont {A.}~\bibnamefont
  {Raccanelli}}, \bibinfo {author} {\bibfnamefont {A.}~\bibnamefont {Bonaldi}},
  \bibinfo {author} {\bibfnamefont {M.}~\bibnamefont {Negrello}}, \bibinfo
  {author} {\bibfnamefont {S.}~\bibnamefont {Matarrese}}, \bibinfo {author}
  {\bibfnamefont {G.}~\bibnamefont {Tormen}}, \ and\ \bibinfo {author}
  {\bibfnamefont {G.~D.}\ \bibnamefont {Zotti}},\ }\href
  {http://arxiv.org/abs/0802.0084} {\bibfield  {journal} {\bibinfo  {journal}
  {Mon.Not.Roy.Astron.Soc.}\ }\textbf {\bibinfo {volume} {386}},\ \bibinfo
  {pages} {2161} (\bibinfo {year} {2008})},\ \Eprint
  {http://arxiv.org/abs/0802.0084} {0802.0084} \BibitemShut {NoStop}%
\bibitem [{\citenamefont {Pullen}\ \emph {et~al.}(2012)\citenamefont {Pullen},
  \citenamefont {Chang}, \citenamefont {Dore},\ and\ \citenamefont
  {Lidz}}]{Pullen:2012}%
  \BibitemOpen
  \bibfield  {author} {\bibinfo {author} {\bibfnamefont {A.}~\bibnamefont
  {Pullen}}, \bibinfo {author} {\bibfnamefont {T.-C.}\ \bibnamefont {Chang}},
  \bibinfo {author} {\bibfnamefont {O.}~\bibnamefont {Dore}}, \ and\ \bibinfo
  {author} {\bibfnamefont {A.}~\bibnamefont {Lidz}},\ }\href
  {http://arxiv.org/abs/1211.1397} {\bibfield  {journal} {\bibinfo  {journal}
  {The Astrophysical Journal, Volume 768, Issue 1, article id. 15, 15 pp.
  (2013)}\ } (\bibinfo {year} {2012})},\ \Eprint
  {http://arxiv.org/abs/1211.1397} {1211.1397} \BibitemShut {NoStop}%
\bibitem [{\citenamefont {Cabre}\ \emph {et~al.}()\citenamefont {Cabre},
  \citenamefont {Fosalba}, \citenamefont {Gaztanaga},\ and\ \citenamefont
  {Manera}}]{Cabre:2007}%
  \BibitemOpen
  \bibfield  {author} {\bibinfo {author} {\bibfnamefont {A.}~\bibnamefont
  {Cabre}}, \bibinfo {author} {\bibfnamefont {P.}~\bibnamefont {Fosalba}},
  \bibinfo {author} {\bibfnamefont {E.}~\bibnamefont {Gaztanaga}}, \ and\
  \bibinfo {author} {\bibfnamefont {M.}~\bibnamefont {Manera}},\ }\href
  {http://arxiv.org/abs/astro-ph/0701393} {\ }\Eprint
  {http://arxiv.org/abs/astro-ph/0701393} {astro-ph/0701393} \BibitemShut
  {NoStop}%
\bibitem [{\citenamefont {Dio}\ \emph {et~al.}(2014)\citenamefont {Dio},
  \citenamefont {Montanari}, \citenamefont {Durrer},\ and\ \citenamefont
  {Lesgourgues}}]{DiDio:2014}%
  \BibitemOpen
  \bibfield  {author} {\bibinfo {author} {\bibfnamefont {E.~D.}\ \bibnamefont
  {Dio}}, \bibinfo {author} {\bibfnamefont {F.}~\bibnamefont {Montanari}},
  \bibinfo {author} {\bibfnamefont {R.}~\bibnamefont {Durrer}}, \ and\ \bibinfo
  {author} {\bibfnamefont {J.}~\bibnamefont {Lesgourgues}},\ }\href
  {http://arxiv.org/abs/1308.6186} {\bibfield  {journal} {\bibinfo  {journal}
  {JCAP}\ }\textbf {\bibinfo {volume} {01}},\ \bibinfo {pages} {042} (\bibinfo
  {year} {2014})},\ \Eprint {http://arxiv.org/abs/1308.6186} {1308.6186}
  \BibitemShut {NoStop}%
\bibitem [{\citenamefont {Vallinotto}(2013)}]{Vallinotto:2013}%
  \BibitemOpen
  \bibfield  {author} {\bibinfo {author} {\bibfnamefont {A.}~\bibnamefont
  {Vallinotto}},\ }\href {http://arxiv.org/abs/1304.3474} {\bibfield  {journal}
  {\bibinfo  {journal} {Astrophys.J.}\ }\textbf {\bibinfo {volume} {778}},\
  \bibinfo {pages} {108} (\bibinfo {year} {2013})},\ \Eprint
  {http://arxiv.org/abs/1304.3474} {1304.3474} \BibitemShut {NoStop}%
\bibitem [{\citenamefont {Giannantonio}\ and\ \citenamefont
  {Percival}(2014)}]{Giannantonio:2014}%
  \BibitemOpen
  \bibfield  {author} {\bibinfo {author} {\bibfnamefont {T.}~\bibnamefont
  {Giannantonio}}\ and\ \bibinfo {author} {\bibfnamefont {W.~J.}\ \bibnamefont
  {Percival}},\ }\href {http://arxiv.org/abs/1312.5154} {\bibfield  {journal}
  {\bibinfo  {journal} {MNRAS Letters, 441,}\ }\textbf {\bibinfo {volume}
  {1}},\ \bibinfo {pages} {L16} (\bibinfo {year} {2014})},\ \Eprint
  {http://arxiv.org/abs/1312.5154} {1312.5154} \BibitemShut {NoStop}%
\bibitem [{\citenamefont {Pujol}\ \emph {et~al.}(2016)\citenamefont {Pujol},
  \citenamefont {Chang}, \citenamefont {Gazta{\~n}aga}, \citenamefont {Amara},
  \citenamefont {Refregier}, \citenamefont {Bacon}, \citenamefont {Carretero},
  \citenamefont {Castander}, \citenamefont {Crocce}, \citenamefont {Fosalba},
  \citenamefont {Manera},\ and\ \citenamefont {Vikram}}]{Pujol:2016}%
  \BibitemOpen
  \bibfield  {author} {\bibinfo {author} {\bibfnamefont {A.}~\bibnamefont
  {Pujol}}, \bibinfo {author} {\bibfnamefont {C.}~\bibnamefont {Chang}},
  \bibinfo {author} {\bibfnamefont {E.}~\bibnamefont {Gazta{\~n}aga}}, \bibinfo
  {author} {\bibfnamefont {A.}~\bibnamefont {Amara}}, \bibinfo {author}
  {\bibfnamefont {A.}~\bibnamefont {Refregier}}, \bibinfo {author}
  {\bibfnamefont {D.~J.}\ \bibnamefont {Bacon}}, \bibinfo {author}
  {\bibfnamefont {J.}~\bibnamefont {Carretero}}, \bibinfo {author}
  {\bibfnamefont {F.~J.}\ \bibnamefont {Castander}}, \bibinfo {author}
  {\bibfnamefont {M.}~\bibnamefont {Crocce}}, \bibinfo {author} {\bibfnamefont
  {P.}~\bibnamefont {Fosalba}}, \bibinfo {author} {\bibfnamefont
  {M.}~\bibnamefont {Manera}}, \ and\ \bibinfo {author} {\bibfnamefont
  {V.}~\bibnamefont {Vikram}},\ }\href {http://arxiv.org/abs/1601.00160} {\
  (\bibinfo {year} {2016})},\ \Eprint {http://arxiv.org/abs/1601.00160}
  {1601.00160} \BibitemShut {NoStop}%
\bibitem [{\citenamefont {Chang}\ \emph {et~al.}(2016)\citenamefont {Chang},
  \citenamefont {Pujol}, \citenamefont {Gaztanaga}, \citenamefont {Amara},
  \citenamefont {Refregier}, \citenamefont {Bacon}, \citenamefont {Becker},
  \citenamefont {Bonnett}, \citenamefont {Carretero}, \citenamefont
  {Castander}, \citenamefont {Crocce}, \citenamefont {Fosalba}, \citenamefont
  {Giannantonio}, \citenamefont {Hartley}, \citenamefont {Jarvis},
  \citenamefont {Kacprzak}, \citenamefont {Ross}, \citenamefont {Sheldon},
  \citenamefont {Troxel}, \citenamefont {Vikram}, \citenamefont {Zuntz},
  \citenamefont {Abbott}, \citenamefont {Abdalla}, \citenamefont {Allam},
  \citenamefont {Annis}, \citenamefont {Benoit-Levy}, \citenamefont {Bertin},
  \citenamefont {Brooks}, \citenamefont {Buckley-Geer}, \citenamefont {Burke},
  \citenamefont {Capozzi}, \citenamefont {Rosell}, \citenamefont {Kind},
  \citenamefont {Cunha}, \citenamefont {D'Andrea}, \citenamefont {da~Costa},
  \citenamefont {Desai}, \citenamefont {Diehl}, \citenamefont {Dietrich},
  \citenamefont {Doel}, \citenamefont {Eifler}, \citenamefont {Estrada},
  \citenamefont {Evrard}, \citenamefont {Flaugher}, \citenamefont {Frieman},
  \citenamefont {Goldstein}, \citenamefont {Gruen}, \citenamefont {Gruendl},
  \citenamefont {Gutierrez}, \citenamefont {Honscheid}, \citenamefont {Jain},
  \citenamefont {James}, \citenamefont {Kuehn}, \citenamefont {Kuropatkin},
  \citenamefont {Lahav}, \citenamefont {Li}, \citenamefont {Lima},
  \citenamefont {Marshall}, \citenamefont {Martini}, \citenamefont {Melchior},
  \citenamefont {Miller},\ and\ \citenamefont {Miquel}}]{Chang:2016}%
  \BibitemOpen
  \bibfield  {author} {\bibinfo {author} {\bibfnamefont {C.}~\bibnamefont
  {Chang}}, \bibinfo {author} {\bibfnamefont {A.}~\bibnamefont {Pujol}},
  \bibinfo {author} {\bibfnamefont {E.}~\bibnamefont {Gaztanaga}}, \bibinfo
  {author} {\bibfnamefont {A.}~\bibnamefont {Amara}}, \bibinfo {author}
  {\bibfnamefont {A.}~\bibnamefont {Refregier}}, \bibinfo {author}
  {\bibfnamefont {D.}~\bibnamefont {Bacon}}, \bibinfo {author} {\bibfnamefont
  {M.~R.}\ \bibnamefont {Becker}}, \bibinfo {author} {\bibfnamefont
  {C.}~\bibnamefont {Bonnett}}, \bibinfo {author} {\bibfnamefont
  {J.}~\bibnamefont {Carretero}}, \bibinfo {author} {\bibfnamefont {F.~J.}\
  \bibnamefont {Castander}}, \bibinfo {author} {\bibfnamefont {M.}~\bibnamefont
  {Crocce}}, \bibinfo {author} {\bibfnamefont {P.}~\bibnamefont {Fosalba}},
  \bibinfo {author} {\bibfnamefont {T.}~\bibnamefont {Giannantonio}}, \bibinfo
  {author} {\bibfnamefont {W.}~\bibnamefont {Hartley}}, \bibinfo {author}
  {\bibfnamefont {M.}~\bibnamefont {Jarvis}}, \bibinfo {author} {\bibfnamefont
  {T.}~\bibnamefont {Kacprzak}}, \bibinfo {author} {\bibfnamefont {A.~J.}\
  \bibnamefont {Ross}}, \bibinfo {author} {\bibfnamefont {E.}~\bibnamefont
  {Sheldon}}, \bibinfo {author} {\bibfnamefont {M.~A.}\ \bibnamefont {Troxel}},
  \bibinfo {author} {\bibfnamefont {V.}~\bibnamefont {Vikram}}, \bibinfo
  {author} {\bibfnamefont {J.}~\bibnamefont {Zuntz}}, \bibinfo {author}
  {\bibfnamefont {T.~M.~C.}\ \bibnamefont {Abbott}}, \bibinfo {author}
  {\bibfnamefont {F.~B.}\ \bibnamefont {Abdalla}}, \bibinfo {author}
  {\bibfnamefont {S.}~\bibnamefont {Allam}}, \bibinfo {author} {\bibfnamefont
  {J.}~\bibnamefont {Annis}}, \bibinfo {author} {\bibfnamefont
  {A.}~\bibnamefont {Benoit-Levy}}, \bibinfo {author} {\bibfnamefont
  {E.}~\bibnamefont {Bertin}}, \bibinfo {author} {\bibfnamefont
  {D.}~\bibnamefont {Brooks}}, \bibinfo {author} {\bibfnamefont
  {E.}~\bibnamefont {Buckley-Geer}}, \bibinfo {author} {\bibfnamefont {D.~L.}\
  \bibnamefont {Burke}}, \bibinfo {author} {\bibfnamefont {D.}~\bibnamefont
  {Capozzi}}, \bibinfo {author} {\bibfnamefont {A.~C.}\ \bibnamefont {Rosell}},
  \bibinfo {author} {\bibfnamefont {M.~C.}\ \bibnamefont {Kind}}, \bibinfo
  {author} {\bibfnamefont {C.~E.}\ \bibnamefont {Cunha}}, \bibinfo {author}
  {\bibfnamefont {C.~B.}\ \bibnamefont {D'Andrea}}, \bibinfo {author}
  {\bibfnamefont {L.~N.}\ \bibnamefont {da~Costa}}, \bibinfo {author}
  {\bibfnamefont {S.}~\bibnamefont {Desai}}, \bibinfo {author} {\bibfnamefont
  {H.~T.}\ \bibnamefont {Diehl}}, \bibinfo {author} {\bibfnamefont {J.~P.}\
  \bibnamefont {Dietrich}}, \bibinfo {author} {\bibfnamefont {P.}~\bibnamefont
  {Doel}}, \bibinfo {author} {\bibfnamefont {T.~F.}\ \bibnamefont {Eifler}},
  \bibinfo {author} {\bibfnamefont {J.}~\bibnamefont {Estrada}}, \bibinfo
  {author} {\bibfnamefont {A.~E.}\ \bibnamefont {Evrard}}, \bibinfo {author}
  {\bibfnamefont {B.}~\bibnamefont {Flaugher}}, \bibinfo {author}
  {\bibfnamefont {J.}~\bibnamefont {Frieman}}, \bibinfo {author} {\bibfnamefont
  {D.~A.}\ \bibnamefont {Goldstein}}, \bibinfo {author} {\bibfnamefont
  {D.}~\bibnamefont {Gruen}}, \bibinfo {author} {\bibfnamefont {R.~A.}\
  \bibnamefont {Gruendl}}, \bibinfo {author} {\bibfnamefont {G.}~\bibnamefont
  {Gutierrez}}, \bibinfo {author} {\bibfnamefont {K.}~\bibnamefont
  {Honscheid}}, \bibinfo {author} {\bibfnamefont {B.}~\bibnamefont {Jain}},
  \bibinfo {author} {\bibfnamefont {D.~J.}\ \bibnamefont {James}}, \bibinfo
  {author} {\bibfnamefont {K.}~\bibnamefont {Kuehn}}, \bibinfo {author}
  {\bibfnamefont {N.}~\bibnamefont {Kuropatkin}}, \bibinfo {author}
  {\bibfnamefont {O.}~\bibnamefont {Lahav}}, \bibinfo {author} {\bibfnamefont
  {T.~S.}\ \bibnamefont {Li}}, \bibinfo {author} {\bibfnamefont
  {M.}~\bibnamefont {Lima}}, \bibinfo {author} {\bibfnamefont {J.~L.}\
  \bibnamefont {Marshall}}, \bibinfo {author} {\bibfnamefont {P.}~\bibnamefont
  {Martini}}, \bibinfo {author} {\bibfnamefont {P.}~\bibnamefont {Melchior}},
  \bibinfo {author} {\bibfnamefont {C.~J.}\ \bibnamefont {Miller}}, \ and\
  \bibinfo {author} {\bibfnamefont {R.}~\bibnamefont {Miquel}},\ }\href
  {http://arxiv.org/abs/1601.00405} {\  (\bibinfo {year} {2016})},\ \Eprint
  {http://arxiv.org/abs/1601.00405} {1601.00405} \BibitemShut {NoStop}%
\bibitem [{\citenamefont {Jarvis}\ \emph {et~al.}(2015)\citenamefont {Jarvis},
  \citenamefont {Bacon}, \citenamefont {Blake}, \citenamefont {Brown},
  \citenamefont {Lindsay}, \citenamefont {Raccanelli}, \citenamefont {Santos},\
  and\ \citenamefont {Schwarz}}]{SKA:Jarvis}%
  \BibitemOpen
  \bibfield  {author} {\bibinfo {author} {\bibfnamefont {M.~J.}\ \bibnamefont
  {Jarvis}}, \bibinfo {author} {\bibfnamefont {D.}~\bibnamefont {Bacon}},
  \bibinfo {author} {\bibfnamefont {C.}~\bibnamefont {Blake}}, \bibinfo
  {author} {\bibfnamefont {M.~L.}\ \bibnamefont {Brown}}, \bibinfo {author}
  {\bibfnamefont {S.~N.}\ \bibnamefont {Lindsay}}, \bibinfo {author}
  {\bibfnamefont {A.}~\bibnamefont {Raccanelli}}, \bibinfo {author}
  {\bibfnamefont {M.}~\bibnamefont {Santos}}, \ and\ \bibinfo {author}
  {\bibfnamefont {D.}~\bibnamefont {Schwarz}},\ }\href
  {http://arxiv.org/abs/1501.03825} {\  (\bibinfo {year} {2015})},\ \Eprint
  {http://arxiv.org/abs/1501.03825} {1501.03825} \BibitemShut {NoStop}%
\bibitem [{\citenamefont {Wilman}\ \emph {et~al.}(2008)\citenamefont {Wilman},
  \citenamefont {Miller}, \citenamefont {Jarvis}, \citenamefont {Mauch},
  \citenamefont {Levrier}, \citenamefont {Abdalla}, \citenamefont {Rawlings},
  \citenamefont {Kloeckner}, \citenamefont {Obreschkow}, \citenamefont
  {Olteanu},\ and\ \citenamefont {Young}}]{Wilman:2008}%
  \BibitemOpen
  \bibfield  {author} {\bibinfo {author} {\bibfnamefont {R.~J.}\ \bibnamefont
  {Wilman}}, \bibinfo {author} {\bibfnamefont {L.}~\bibnamefont {Miller}},
  \bibinfo {author} {\bibfnamefont {M.~J.}\ \bibnamefont {Jarvis}}, \bibinfo
  {author} {\bibfnamefont {T.}~\bibnamefont {Mauch}}, \bibinfo {author}
  {\bibfnamefont {F.}~\bibnamefont {Levrier}}, \bibinfo {author} {\bibfnamefont
  {F.~B.}\ \bibnamefont {Abdalla}}, \bibinfo {author} {\bibfnamefont
  {S.}~\bibnamefont {Rawlings}}, \bibinfo {author} {\bibfnamefont {H.-R.}\
  \bibnamefont {Kloeckner}}, \bibinfo {author} {\bibfnamefont {D.}~\bibnamefont
  {Obreschkow}}, \bibinfo {author} {\bibfnamefont {D.}~\bibnamefont {Olteanu}},
  \ and\ \bibinfo {author} {\bibfnamefont {S.}~\bibnamefont {Young}},\ }\href
  {http://arxiv.org/abs/0805.3413} {\bibfield  {journal} {\bibinfo  {journal}
  {Mon.Not.Roy.Astron.Soc.}\ }\textbf {\bibinfo {volume} {388}},\ \bibinfo
  {pages} {1335} (\bibinfo {year} {2008})},\ \Eprint
  {http://arxiv.org/abs/0805.3413} {0805.3413} \BibitemShut {NoStop}%
\bibitem [{\citenamefont {M{\'e}nard}\ \emph {et~al.}(2013)\citenamefont
  {M{\'e}nard}, \citenamefont {Scranton}, \citenamefont {Schmidt},
  \citenamefont {Morrison}, \citenamefont {Jeong}, \citenamefont {Budavari},\
  and\ \citenamefont {Rahman}}]{Menard:2013}%
  \BibitemOpen
  \bibfield  {author} {\bibinfo {author} {\bibfnamefont {B.}~\bibnamefont
  {M{\'e}nard}}, \bibinfo {author} {\bibfnamefont {R.}~\bibnamefont
  {Scranton}}, \bibinfo {author} {\bibfnamefont {S.}~\bibnamefont {Schmidt}},
  \bibinfo {author} {\bibfnamefont {C.}~\bibnamefont {Morrison}}, \bibinfo
  {author} {\bibfnamefont {D.}~\bibnamefont {Jeong}}, \bibinfo {author}
  {\bibfnamefont {T.}~\bibnamefont {Budavari}}, \ and\ \bibinfo {author}
  {\bibfnamefont {M.}~\bibnamefont {Rahman}},\ }\href
  {http://arxiv.org/abs/1303.4722} {\  (\bibinfo {year} {2013})},\ \Eprint
  {http://arxiv.org/abs/1303.4722} {1303.4722} \BibitemShut {NoStop}%
\bibitem [{\citenamefont {Rahman}\ \emph {et~al.}(2014)\citenamefont {Rahman},
  \citenamefont {M{\'e}nard}, \citenamefont {Scranton}, \citenamefont
  {Schmidt},\ and\ \citenamefont {Morrison}}]{Rahman:2014}%
  \BibitemOpen
  \bibfield  {author} {\bibinfo {author} {\bibfnamefont {M.}~\bibnamefont
  {Rahman}}, \bibinfo {author} {\bibfnamefont {B.}~\bibnamefont {M{\'e}nard}},
  \bibinfo {author} {\bibfnamefont {R.}~\bibnamefont {Scranton}}, \bibinfo
  {author} {\bibfnamefont {S.~J.}\ \bibnamefont {Schmidt}}, \ and\ \bibinfo
  {author} {\bibfnamefont {C.~B.}\ \bibnamefont {Morrison}},\ }\href
  {http://arxiv.org/abs/1407.7860} {\  (\bibinfo {year} {2014})},\ \Eprint
  {http://arxiv.org/abs/1407.7860} {1407.7860} \BibitemShut {NoStop}%
\bibitem [{\citenamefont {{Kovetz}}\ \emph {et~al.}(tion)\citenamefont
  {{Kovetz}}, \citenamefont {{Raccanelli}},\ and\ \citenamefont
  {{Rahman}}}]{Kovetz:CBR}%
  \BibitemOpen
  \bibfield  {author} {\bibinfo {author} {\bibfnamefont {E.~D.}\ \bibnamefont
  {{Kovetz}}}, \bibinfo {author} {\bibfnamefont {A.}~\bibnamefont
  {{Raccanelli}}}, \ and\ \bibinfo {author} {\bibfnamefont {M.}~\bibnamefont
  {{Rahman}}},\ }\href@noop {} {\  (\bibinfo {year} {in
  preparation})}\BibitemShut {NoStop}%
\bibitem [{\citenamefont {Schutz}(2011)}]{Schutz:2011}%
  \BibitemOpen
  \bibfield  {author} {\bibinfo {author} {\bibfnamefont {B.~F.}\ \bibnamefont
  {Schutz}},\ }\href {http://arxiv.org/abs/1102.5421} {\  (\bibinfo {year}
  {2011})},\ \Eprint {http://arxiv.org/abs/1102.5421} {1102.5421} \BibitemShut
  {NoStop}%
\bibitem [{\citenamefont {Klimenko}\ \emph {et~al.}(2011)\citenamefont
  {Klimenko}, \citenamefont {Vedovato}, \citenamefont {Drago}, \citenamefont
  {Mazzolo}, \citenamefont {Mitselmakher}, \citenamefont {Pankow},
  \citenamefont {Prodi}, \citenamefont {Re}, \citenamefont {Salemi},\ and\
  \citenamefont {Yakushin}}]{Klimenko:2011}%
  \BibitemOpen
  \bibfield  {author} {\bibinfo {author} {\bibfnamefont {S.}~\bibnamefont
  {Klimenko}}, \bibinfo {author} {\bibfnamefont {G.}~\bibnamefont {Vedovato}},
  \bibinfo {author} {\bibfnamefont {M.}~\bibnamefont {Drago}}, \bibinfo
  {author} {\bibfnamefont {G.}~\bibnamefont {Mazzolo}}, \bibinfo {author}
  {\bibfnamefont {G.}~\bibnamefont {Mitselmakher}}, \bibinfo {author}
  {\bibfnamefont {C.}~\bibnamefont {Pankow}}, \bibinfo {author} {\bibfnamefont
  {G.}~\bibnamefont {Prodi}}, \bibinfo {author} {\bibfnamefont
  {V.}~\bibnamefont {Re}}, \bibinfo {author} {\bibfnamefont {F.}~\bibnamefont
  {Salemi}}, \ and\ \bibinfo {author} {\bibfnamefont {I.}~\bibnamefont
  {Yakushin}},\ }\href {http://arxiv.org/abs/1101.5408} {\bibfield  {journal}
  {\bibinfo  {journal} {Phys.Rev.D}\ }\textbf {\bibinfo {volume} {83}},\
  \bibinfo {pages} {102001} (\bibinfo {year} {2011})},\ \Eprint
  {http://arxiv.org/abs/1101.5408} {1101.5408} \BibitemShut {NoStop}%
\bibitem [{\citenamefont {Sidery}\ \emph {et~al.}(2014)\citenamefont {Sidery},
  \citenamefont {Aylott}, \citenamefont {Christensen}, \citenamefont {Farr},
  \citenamefont {Farr}, \citenamefont {Feroz}, \citenamefont {Gair},
  \citenamefont {Grover}, \citenamefont {Graff}, \citenamefont {Hanna},
  \citenamefont {Kalogera}, \citenamefont {Mandel}, \citenamefont
  {O'Shaughnessy}, \citenamefont {Pitkin}, \citenamefont {Price}, \citenamefont
  {Raymond}, \citenamefont {Roever}, \citenamefont {Singer}, \citenamefont {der
  Sluys}, \citenamefont {Smith}, \citenamefont {Vecchio}, \citenamefont
  {Veitch},\ and\ \citenamefont {Vitale}}]{Sidery:2014}%
  \BibitemOpen
  \bibfield  {author} {\bibinfo {author} {\bibfnamefont {T.}~\bibnamefont
  {Sidery}}, \bibinfo {author} {\bibfnamefont {B.}~\bibnamefont {Aylott}},
  \bibinfo {author} {\bibfnamefont {N.}~\bibnamefont {Christensen}}, \bibinfo
  {author} {\bibfnamefont {B.}~\bibnamefont {Farr}}, \bibinfo {author}
  {\bibfnamefont {W.}~\bibnamefont {Farr}}, \bibinfo {author} {\bibfnamefont
  {F.}~\bibnamefont {Feroz}}, \bibinfo {author} {\bibfnamefont
  {J.}~\bibnamefont {Gair}}, \bibinfo {author} {\bibfnamefont {K.}~\bibnamefont
  {Grover}}, \bibinfo {author} {\bibfnamefont {P.}~\bibnamefont {Graff}},
  \bibinfo {author} {\bibfnamefont {C.}~\bibnamefont {Hanna}}, \bibinfo
  {author} {\bibfnamefont {V.}~\bibnamefont {Kalogera}}, \bibinfo {author}
  {\bibfnamefont {I.}~\bibnamefont {Mandel}}, \bibinfo {author} {\bibfnamefont
  {R.}~\bibnamefont {O'Shaughnessy}}, \bibinfo {author} {\bibfnamefont
  {M.}~\bibnamefont {Pitkin}}, \bibinfo {author} {\bibfnamefont
  {L.}~\bibnamefont {Price}}, \bibinfo {author} {\bibfnamefont
  {V.}~\bibnamefont {Raymond}}, \bibinfo {author} {\bibfnamefont
  {C.}~\bibnamefont {Roever}}, \bibinfo {author} {\bibfnamefont
  {L.}~\bibnamefont {Singer}}, \bibinfo {author} {\bibfnamefont {M.~V.}\
  \bibnamefont {der Sluys}}, \bibinfo {author} {\bibfnamefont {R.~J.}\
  \bibnamefont {Smith}}, \bibinfo {author} {\bibfnamefont {A.}~\bibnamefont
  {Vecchio}}, \bibinfo {author} {\bibfnamefont {J.}~\bibnamefont {Veitch}}, \
  and\ \bibinfo {author} {\bibfnamefont {S.}~\bibnamefont {Vitale}},\ }\href
  {http://arxiv.org/abs/1312.6013} {\bibfield  {journal} {\bibinfo  {journal}
  {Phys. Rev. D}\ }\textbf {\bibinfo {volume} {89}},\ \bibinfo {pages} {084060}
  (\bibinfo {year} {2014})},\ \Eprint {http://arxiv.org/abs/1312.6013}
  {1312.6013} \BibitemShut {NoStop}%
\bibitem [{\citenamefont {Namikawa}\ \emph
  {et~al.}(2016{\natexlab{b}})\citenamefont {Namikawa}, \citenamefont
  {Nishizawa},\ and\ \citenamefont {Taruya}}]{Namikawa:2016noz}%
  \BibitemOpen
  \bibfield  {author} {\bibinfo {author} {\bibfnamefont {T.}~\bibnamefont
  {Namikawa}}, \bibinfo {author} {\bibfnamefont {A.}~\bibnamefont {Nishizawa}},
  \ and\ \bibinfo {author} {\bibfnamefont {A.}~\bibnamefont {Taruya}},\ }\href
  {http://arxiv.org/abs/1511.04638} {\bibfield  {journal} {\bibinfo  {journal}
  {Phys. Rev. Lett.}\ }\textbf {\bibinfo {volume} {116}},\ \bibinfo {pages}
  {121302} (\bibinfo {year} {2016}{\natexlab{b}})},\ \Eprint
  {http://arxiv.org/abs/1511.04638} {1511.04638} \BibitemShut {NoStop}%
\bibitem [{\citenamefont {Finn}\ \emph {et~al.}(2010)\citenamefont {Finn},
  \citenamefont {Fritschel}, \citenamefont {Klimenko}, \citenamefont {Raab},
  \citenamefont {Sathyaprakash}, \citenamefont {Saulson},\ and\ \citenamefont
  {Weiss}}]{LIGO-net}%
  \BibitemOpen
  \bibfield  {author} {\bibinfo {author} {\bibfnamefont {S.}~\bibnamefont
  {Finn}}, \bibinfo {author} {\bibfnamefont {P.}~\bibnamefont {Fritschel}},
  \bibinfo {author} {\bibfnamefont {S.}~\bibnamefont {Klimenko}}, \bibinfo
  {author} {\bibfnamefont {F.}~\bibnamefont {Raab}}, \bibinfo {author}
  {\bibfnamefont {B.}~\bibnamefont {Sathyaprakash}}, \bibinfo {author}
  {\bibfnamefont {P.}~\bibnamefont {Saulson}}, \ and\ \bibinfo {author}
  {\bibfnamefont {R.}~\bibnamefont {Weiss}},\ }\href@noop {} {\  (\bibinfo
  {year} {2010})}\BibitemShut {NoStop}%
\bibitem [{\citenamefont {Abbott}\ \emph {et~al.}(2016)\citenamefont {Abbott},
  \citenamefont {Abbott}, \citenamefont {Abbott}, \citenamefont {Abernathy},
  \citenamefont {Acernese}, \citenamefont {Ackley}, \citenamefont {Adams},
  \citenamefont {Adams}, \citenamefont {Addesso}, \citenamefont {Adhikari},
  \citenamefont {Adya}, \citenamefont {Affeldt}, \citenamefont {Agathos},
  \citenamefont {Agatsuma}, \citenamefont {Aggarwal}, \citenamefont {Aguiar},
  \citenamefont {Aiello}, \citenamefont {Ain}, \citenamefont {Ajith},
  \citenamefont {Allen}, \citenamefont {Allocca}, \citenamefont {Altin},
  \citenamefont {Anderson}, \citenamefont {Anderson}, \citenamefont {Arai},
  \citenamefont {Araya}, \citenamefont {Arceneaux}, \citenamefont {Areeda},
  \citenamefont {Arnaud}, \citenamefont {Arun}, \citenamefont {Ascenzi},
  \citenamefont {Ashton}, \citenamefont {Ast}, \citenamefont {Aston},
  \citenamefont {Astone}, \citenamefont {Aufmuth}, \citenamefont {Aulbert},
  \citenamefont {Babak}, \citenamefont {Bacon}, \citenamefont {Bader},
  \citenamefont {Baker}, \citenamefont {Baldaccini}, \citenamefont {Ballardin},
  \citenamefont {Ballmer}, \citenamefont {Barayoga}, \citenamefont {Barclay},
  \citenamefont {Barish}, \citenamefont {Barker}, \citenamefont {Barone},
  \citenamefont {Barr}, \citenamefont {Barsotti}, \citenamefont {Barsuglia},
  \citenamefont {Barta}, \citenamefont {Bartlett}, \citenamefont {Bartos},
  \citenamefont {Bassiri}, \citenamefont {Basti}, \citenamefont {Batch},
  \citenamefont {Baune}, \citenamefont {Bavigadda}, \citenamefont {Bazzan},
  \citenamefont {Behnke}, \citenamefont {Bejger}, \citenamefont {Belczynski},\
  and\ \citenamefont {Bell}}]{LIGO:rate}%
  \BibitemOpen
  \bibfield  {author} {\bibinfo {author} {\bibfnamefont {B.~P.}\ \bibnamefont
  {Abbott}}, \bibinfo {author} {\bibfnamefont {R.}~\bibnamefont {Abbott}},
  \bibinfo {author} {\bibfnamefont {T.~D.}\ \bibnamefont {Abbott}}, \bibinfo
  {author} {\bibfnamefont {M.~R.}\ \bibnamefont {Abernathy}}, \bibinfo {author}
  {\bibfnamefont {F.}~\bibnamefont {Acernese}}, \bibinfo {author}
  {\bibfnamefont {K.}~\bibnamefont {Ackley}}, \bibinfo {author} {\bibfnamefont
  {C.}~\bibnamefont {Adams}}, \bibinfo {author} {\bibfnamefont
  {T.}~\bibnamefont {Adams}}, \bibinfo {author} {\bibfnamefont
  {P.}~\bibnamefont {Addesso}}, \bibinfo {author} {\bibfnamefont {R.~X.}\
  \bibnamefont {Adhikari}}, \bibinfo {author} {\bibfnamefont {V.~B.}\
  \bibnamefont {Adya}}, \bibinfo {author} {\bibfnamefont {C.}~\bibnamefont
  {Affeldt}}, \bibinfo {author} {\bibfnamefont {M.}~\bibnamefont {Agathos}},
  \bibinfo {author} {\bibfnamefont {K.}~\bibnamefont {Agatsuma}}, \bibinfo
  {author} {\bibfnamefont {N.}~\bibnamefont {Aggarwal}}, \bibinfo {author}
  {\bibfnamefont {O.~D.}\ \bibnamefont {Aguiar}}, \bibinfo {author}
  {\bibfnamefont {L.}~\bibnamefont {Aiello}}, \bibinfo {author} {\bibfnamefont
  {A.}~\bibnamefont {Ain}}, \bibinfo {author} {\bibfnamefont {P.}~\bibnamefont
  {Ajith}}, \bibinfo {author} {\bibfnamefont {B.}~\bibnamefont {Allen}},
  \bibinfo {author} {\bibfnamefont {A.}~\bibnamefont {Allocca}}, \bibinfo
  {author} {\bibfnamefont {P.~A.}\ \bibnamefont {Altin}}, \bibinfo {author}
  {\bibfnamefont {S.~B.}\ \bibnamefont {Anderson}}, \bibinfo {author}
  {\bibfnamefont {W.~G.}\ \bibnamefont {Anderson}}, \bibinfo {author}
  {\bibfnamefont {K.}~\bibnamefont {Arai}}, \bibinfo {author} {\bibfnamefont
  {M.~C.}\ \bibnamefont {Araya}}, \bibinfo {author} {\bibfnamefont {C.~C.}\
  \bibnamefont {Arceneaux}}, \bibinfo {author} {\bibfnamefont {J.~S.}\
  \bibnamefont {Areeda}}, \bibinfo {author} {\bibfnamefont {N.}~\bibnamefont
  {Arnaud}}, \bibinfo {author} {\bibfnamefont {K.~G.}\ \bibnamefont {Arun}},
  \bibinfo {author} {\bibfnamefont {S.}~\bibnamefont {Ascenzi}}, \bibinfo
  {author} {\bibfnamefont {G.}~\bibnamefont {Ashton}}, \bibinfo {author}
  {\bibfnamefont {M.}~\bibnamefont {Ast}}, \bibinfo {author} {\bibfnamefont
  {S.~M.}\ \bibnamefont {Aston}}, \bibinfo {author} {\bibfnamefont
  {P.}~\bibnamefont {Astone}}, \bibinfo {author} {\bibfnamefont
  {P.}~\bibnamefont {Aufmuth}}, \bibinfo {author} {\bibfnamefont
  {C.}~\bibnamefont {Aulbert}}, \bibinfo {author} {\bibfnamefont
  {S.}~\bibnamefont {Babak}}, \bibinfo {author} {\bibfnamefont
  {P.}~\bibnamefont {Bacon}}, \bibinfo {author} {\bibfnamefont {M.~K.~M.}\
  \bibnamefont {Bader}}, \bibinfo {author} {\bibfnamefont {P.~T.}\ \bibnamefont
  {Baker}}, \bibinfo {author} {\bibfnamefont {F.}~\bibnamefont {Baldaccini}},
  \bibinfo {author} {\bibfnamefont {G.}~\bibnamefont {Ballardin}}, \bibinfo
  {author} {\bibfnamefont {S.~W.}\ \bibnamefont {Ballmer}}, \bibinfo {author}
  {\bibfnamefont {J.~C.}\ \bibnamefont {Barayoga}}, \bibinfo {author}
  {\bibfnamefont {S.~E.}\ \bibnamefont {Barclay}}, \bibinfo {author}
  {\bibfnamefont {B.~C.}\ \bibnamefont {Barish}}, \bibinfo {author}
  {\bibfnamefont {D.}~\bibnamefont {Barker}}, \bibinfo {author} {\bibfnamefont
  {F.}~\bibnamefont {Barone}}, \bibinfo {author} {\bibfnamefont
  {B.}~\bibnamefont {Barr}}, \bibinfo {author} {\bibfnamefont {L.}~\bibnamefont
  {Barsotti}}, \bibinfo {author} {\bibfnamefont {M.}~\bibnamefont {Barsuglia}},
  \bibinfo {author} {\bibfnamefont {D.}~\bibnamefont {Barta}}, \bibinfo
  {author} {\bibfnamefont {J.}~\bibnamefont {Bartlett}}, \bibinfo {author}
  {\bibfnamefont {I.}~\bibnamefont {Bartos}}, \bibinfo {author} {\bibfnamefont
  {R.}~\bibnamefont {Bassiri}}, \bibinfo {author} {\bibfnamefont
  {A.}~\bibnamefont {Basti}}, \bibinfo {author} {\bibfnamefont {J.~C.}\
  \bibnamefont {Batch}}, \bibinfo {author} {\bibfnamefont {C.}~\bibnamefont
  {Baune}}, \bibinfo {author} {\bibfnamefont {V.}~\bibnamefont {Bavigadda}},
  \bibinfo {author} {\bibfnamefont {M.}~\bibnamefont {Bazzan}}, \bibinfo
  {author} {\bibfnamefont {B.}~\bibnamefont {Behnke}}, \bibinfo {author}
  {\bibfnamefont {M.}~\bibnamefont {Bejger}}, \bibinfo {author} {\bibfnamefont
  {C.}~\bibnamefont {Belczynski}}, \ and\ \bibinfo {author} {\bibfnamefont
  {A.~S.}\ \bibnamefont {Bell}},\ }\href {http://arxiv.org/abs/1602.03842} {\
  (\bibinfo {year} {2016})},\ \Eprint {http://arxiv.org/abs/1602.03842}
  {1602.03842} \BibitemShut {NoStop}%
\bibitem [{\citenamefont {Dominik}\ \emph {et~al.}(2013)\citenamefont
  {Dominik}, \citenamefont {Belczynski}, \citenamefont {Fryer}, \citenamefont
  {Holz}, \citenamefont {Berti}, \citenamefont {Bulik}, \citenamefont
  {Mandel},\ and\ \citenamefont {O'Shaughnessy}}]{Dominik:2013}%
  \BibitemOpen
  \bibfield  {author} {\bibinfo {author} {\bibfnamefont {M.}~\bibnamefont
  {Dominik}}, \bibinfo {author} {\bibfnamefont {K.}~\bibnamefont {Belczynski}},
  \bibinfo {author} {\bibfnamefont {C.}~\bibnamefont {Fryer}}, \bibinfo
  {author} {\bibfnamefont {D.~E.}\ \bibnamefont {Holz}}, \bibinfo {author}
  {\bibfnamefont {E.}~\bibnamefont {Berti}}, \bibinfo {author} {\bibfnamefont
  {T.}~\bibnamefont {Bulik}}, \bibinfo {author} {\bibfnamefont
  {I.}~\bibnamefont {Mandel}}, \ and\ \bibinfo {author} {\bibfnamefont
  {R.}~\bibnamefont {O'Shaughnessy}},\ }\href {http://arxiv.org/abs/1308.1546}
  {\  (\bibinfo {year} {2013})},\ \Eprint {http://arxiv.org/abs/1308.1546}
  {1308.1546} \BibitemShut {NoStop}%
\bibitem [{\citenamefont {Matsubara}()}]{Matsubara:2000}%
  \BibitemOpen
  \bibfield  {author} {\bibinfo {author} {\bibfnamefont {T.}~\bibnamefont
  {Matsubara}},\ }\href {http://arxiv.org/abs/astro-ph/0004392} {\ }\Eprint
  {http://arxiv.org/abs/astro-ph/0004392} {astro-ph/0004392} \BibitemShut
  {NoStop}%
\bibitem [{\citenamefont {Camera}\ and\ \citenamefont
  {Nishizawa}(2013)}]{Camera:2013}%
  \BibitemOpen
  \bibfield  {author} {\bibinfo {author} {\bibfnamefont {S.}~\bibnamefont
  {Camera}}\ and\ \bibinfo {author} {\bibfnamefont {A.}~\bibnamefont
  {Nishizawa}},\ }\href {http://arxiv.org/abs/1303.5446} {\bibfield  {journal}
  {\bibinfo  {journal} {PRL}\ }\textbf {\bibinfo {volume} {110}},\ \bibinfo
  {pages} {151103} (\bibinfo {year} {2013})},\ \Eprint
  {http://arxiv.org/abs/1303.5446} {1303.5446} \BibitemShut {NoStop}%
\bibitem [{\citenamefont {Oguri}(2016)}]{Oguri:2016}%
  \BibitemOpen
  \bibfield  {author} {\bibinfo {author} {\bibfnamefont {M.}~\bibnamefont
  {Oguri}},\ }\href {http://arxiv.org/abs/1603.02356} {\  (\bibinfo {year}
  {2016})},\ \Eprint {http://arxiv.org/abs/1603.02356} {1603.02356}
  \BibitemShut {NoStop}%
\bibitem [{\citenamefont {Mo}\ and\ \citenamefont {White}()}]{Mo:1995}%
  \BibitemOpen
  \bibfield  {author} {\bibinfo {author} {\bibfnamefont {H.~J.}\ \bibnamefont
  {Mo}}\ and\ \bibinfo {author} {\bibfnamefont {S.~D.~M.}\ \bibnamefont
  {White}},\ }\href {http://arxiv.org/abs/astro-ph/9412088} {\ }\Eprint
  {http://arxiv.org/abs/astro-ph/9412088} {astro-ph/9412088} \BibitemShut
  {NoStop}%
\bibitem [{\citenamefont {Tegmark}\ and\ \citenamefont
  {Peebles}(1998)}]{Tegmark:1998}%
  \BibitemOpen
  \bibfield  {author} {\bibinfo {author} {\bibfnamefont {M.}~\bibnamefont
  {Tegmark}}\ and\ \bibinfo {author} {\bibfnamefont {P.~J.~E.}\ \bibnamefont
  {Peebles}},\ }\href {http://arxiv.org/abs/astro-ph/9804067} {\bibfield
  {journal} {\bibinfo  {journal} {ApJL,}\ }\textbf {\bibinfo {volume} {500}},\
  \bibinfo {pages} {79} (\bibinfo {year} {1998})},\ \Eprint
  {http://arxiv.org/abs/astro-ph/9804067} {astro-ph/9804067} \BibitemShut
  {NoStop}%
\bibitem [{\citenamefont {Jeong}\ and\ \citenamefont
  {Kamionkowski}(2012)}]{Jeong:2012fossils}%
  \BibitemOpen
  \bibfield  {author} {\bibinfo {author} {\bibfnamefont {D.}~\bibnamefont
  {Jeong}}\ and\ \bibinfo {author} {\bibfnamefont {M.}~\bibnamefont
  {Kamionkowski}},\ }\href {http://arxiv.org/abs/1203.0302} {\  (\bibinfo
  {year} {2012})},\ \Eprint {http://arxiv.org/abs/1203.0302} {1203.0302}
  \BibitemShut {NoStop}%
\bibitem [{\citenamefont {Dai}\ \emph {et~al.}(2016)\citenamefont {Dai},
  \citenamefont {Kamionkowski}, \citenamefont {Kovetz}, \citenamefont
  {Raccanelli},\ and\ \citenamefont {Shiraishi}}]{Dai:2016}%
  \BibitemOpen
  \bibfield  {author} {\bibinfo {author} {\bibfnamefont {L.}~\bibnamefont
  {Dai}}, \bibinfo {author} {\bibfnamefont {M.}~\bibnamefont {Kamionkowski}},
  \bibinfo {author} {\bibfnamefont {E.~D.}\ \bibnamefont {Kovetz}}, \bibinfo
  {author} {\bibfnamefont {A.}~\bibnamefont {Raccanelli}}, \ and\ \bibinfo
  {author} {\bibfnamefont {M.}~\bibnamefont {Shiraishi}},\ }\href
  {http://arxiv.org/abs/1507.05618} {\bibfield  {journal} {\bibinfo  {journal}
  {Phys. Rev. D}\ }\textbf {\bibinfo {volume} {93}},\ \bibinfo {pages} {023507}
  (\bibinfo {year} {2016})},\ \Eprint {http://arxiv.org/abs/1507.05618}
  {1507.05618} \BibitemShut {NoStop}%
\bibitem [{\citenamefont {Yoo}(2010)}]{Yoo:2010}%
  \BibitemOpen
  \bibfield  {author} {\bibinfo {author} {\bibfnamefont {J.}~\bibnamefont
  {Yoo}},\ }\href@noop {} {\bibfield  {journal} {\bibinfo  {journal}
  {Phys.Rev.D}\ }\textbf {\bibinfo {volume} {82}},\ \bibinfo {pages} {083508}
  (\bibinfo {year} {2010})},\ \Eprint {http://arxiv.org/abs/1009.3021}
  {1009.3021} \BibitemShut {NoStop}%
\bibitem [{\citenamefont {Bonvin}\ and\ \citenamefont
  {Durrer}(2011)}]{Bonvin:2011}%
  \BibitemOpen
  \bibfield  {author} {\bibinfo {author} {\bibfnamefont {C.}~\bibnamefont
  {Bonvin}}\ and\ \bibinfo {author} {\bibfnamefont {R.}~\bibnamefont
  {Durrer}},\ }\href@noop {} {\bibfield  {journal} {\bibinfo  {journal}
  {Phys.Rev.D}\ }\textbf {\bibinfo {volume} {84}},\ \bibinfo {pages} {063505}
  (\bibinfo {year} {2011})},\ \Eprint {http://arxiv.org/abs/1105.5280}
  {1105.5280} \BibitemShut {NoStop}%
\bibitem [{\citenamefont {Challinor}\ and\ \citenamefont
  {Lewis}(2011)}]{Challinor:2011}%
  \BibitemOpen
  \bibfield  {author} {\bibinfo {author} {\bibfnamefont {A.}~\bibnamefont
  {Challinor}}\ and\ \bibinfo {author} {\bibfnamefont {A.}~\bibnamefont
  {Lewis}},\ }\href {http://arxiv.org/abs/1105.5292} {\bibfield  {journal}
  {\bibinfo  {journal} {Phys.Rev.D}\ }\textbf {\bibinfo {volume} {84}},\
  \bibinfo {pages} {043516} (\bibinfo {year} {2011})},\ \Eprint
  {http://arxiv.org/abs/1105.5292} {1105.5292} \BibitemShut {NoStop}%
\bibitem [{\citenamefont {Yoo}\ \emph {et~al.}(2012)\citenamefont {Yoo},
  \citenamefont {Hamaus}, \citenamefont {Seljak},\ and\ \citenamefont
  {Zaldarriaga}}]{Yoo:2012}%
  \BibitemOpen
  \bibfield  {author} {\bibinfo {author} {\bibfnamefont {J.}~\bibnamefont
  {Yoo}}, \bibinfo {author} {\bibfnamefont {N.}~\bibnamefont {Hamaus}},
  \bibinfo {author} {\bibfnamefont {U.}~\bibnamefont {Seljak}}, \ and\ \bibinfo
  {author} {\bibfnamefont {M.}~\bibnamefont {Zaldarriaga}},\ }\href@noop {}
  {\bibfield  {journal} {\bibinfo  {journal} {Phys.Rev.D}\ }\textbf {\bibinfo
  {volume} {86}},\ \bibinfo {pages} {063514} (\bibinfo {year} {2012})},\
  \Eprint {http://arxiv.org/abs/1109.0998} {1109.0998} \BibitemShut {NoStop}%
\bibitem [{\citenamefont {Jeong}\ \emph {et~al.}(2011)\citenamefont {Jeong},
  \citenamefont {Schmidt},\ and\ \citenamefont {Hirata}}]{Jeong:2012}%
  \BibitemOpen
  \bibfield  {author} {\bibinfo {author} {\bibfnamefont {D.}~\bibnamefont
  {Jeong}}, \bibinfo {author} {\bibfnamefont {F.}~\bibnamefont {Schmidt}}, \
  and\ \bibinfo {author} {\bibfnamefont {C.~M.}\ \bibnamefont {Hirata}},\
  }\href {http://arxiv.org/abs/1107.5427} {\bibfield  {journal} {\bibinfo
  {journal} {2012, PRD 85, 023504}\ } (\bibinfo {year} {2011})},\ \Eprint
  {http://arxiv.org/abs/1107.5427} {1107.5427} \BibitemShut {NoStop}%
\bibitem [{\citenamefont {Bertacca}\ \emph {et~al.}(2012)\citenamefont
  {Bertacca}, \citenamefont {Maartens}, \citenamefont {Raccanelli},\ and\
  \citenamefont {Clarkson}}]{Bertacca:2012}%
  \BibitemOpen
  \bibfield  {author} {\bibinfo {author} {\bibfnamefont {D.}~\bibnamefont
  {Bertacca}}, \bibinfo {author} {\bibfnamefont {R.}~\bibnamefont {Maartens}},
  \bibinfo {author} {\bibfnamefont {A.}~\bibnamefont {Raccanelli}}, \ and\
  \bibinfo {author} {\bibfnamefont {C.}~\bibnamefont {Clarkson}},\ }\href
  {http://arxiv.org/abs/1205.5221} {\bibfield  {journal} {\bibinfo  {journal}
  {JCAP10(2012)025}\ } (\bibinfo {year} {2012})},\ \Eprint
  {http://arxiv.org/abs/1205.5221} {1205.5221} \BibitemShut {NoStop}%
\bibitem [{\citenamefont {Raccanelli}\ \emph {et~al.}(2015)\citenamefont
  {Raccanelli}, \citenamefont {Montanari}, \citenamefont {Bertacca},
  \citenamefont {Dor{\'e}},\ and\ \citenamefont {Durrer}}]{Raccanelli:2015GR}%
  \BibitemOpen
  \bibfield  {author} {\bibinfo {author} {\bibfnamefont {A.}~\bibnamefont
  {Raccanelli}}, \bibinfo {author} {\bibfnamefont {F.}~\bibnamefont
  {Montanari}}, \bibinfo {author} {\bibfnamefont {D.}~\bibnamefont {Bertacca}},
  \bibinfo {author} {\bibfnamefont {O.}~\bibnamefont {Dor{\'e}}}, \ and\
  \bibinfo {author} {\bibfnamefont {R.}~\bibnamefont {Durrer}},\ }\href
  {http://arxiv.org/abs/1505.06179} {\  (\bibinfo {year} {2015})},\ \Eprint
  {http://arxiv.org/abs/1505.06179} {1505.06179} \BibitemShut {NoStop}%
\bibitem [{\citenamefont {Kawamura}\ \emph {et~al.}(2011)\citenamefont
  {Kawamura} \emph {et~al.}}]{decigo}%
  \BibitemOpen
  \bibfield  {author} {\bibinfo {author} {\bibfnamefont {S.}~\bibnamefont
  {Kawamura}} \emph {et~al.},\ }\bibfield  {booktitle} {\emph {\bibinfo
  {booktitle} {{Laser interferometer space antenna. Proceedings, 8th
  International LISA Symposium, Stanford, USA, June 28-July 2, 2010}}},\ }\href
  {\doibase 10.1088/0264-9381/28/9/094011} {\bibfield  {journal} {\bibinfo
  {journal} {Class. Quant. Grav.}\ }\textbf {\bibinfo {volume} {28}},\ \bibinfo
  {pages} {094011} (\bibinfo {year} {2011})}\BibitemShut {NoStop}%
\bibitem [{\citenamefont {Takada}\ \emph {et~al.}(2012)\citenamefont {Takada},
  \citenamefont {Ellis}, \citenamefont {Chiba}, \citenamefont {Greene},
  \citenamefont {Aihara}, \citenamefont {Arimoto}, \citenamefont {Bundy},
  \citenamefont {Cohen}, \citenamefont {Dor{\'e}}, \citenamefont {Graves},
  \citenamefont {Gunn}, \citenamefont {Heckman}, \citenamefont {Hirata},
  \citenamefont {Ho}, \citenamefont {Kneib}, \citenamefont {F{\`e}vre},
  \citenamefont {Lin}, \citenamefont {More}, \citenamefont {Murayama},
  \citenamefont {Nagao}, \citenamefont {Ouchi}, \citenamefont {Seiffert},
  \citenamefont {Silverman}, \citenamefont {Jr}, \citenamefont {Spergel},
  \citenamefont {Strauss}, \citenamefont {Sugai}, \citenamefont {Suto},
  \citenamefont {Takami},\ and\ \citenamefont {Wyse}}]{pfs}%
  \BibitemOpen
  \bibfield  {author} {\bibinfo {author} {\bibfnamefont {M.}~\bibnamefont
  {Takada}}, \bibinfo {author} {\bibfnamefont {R.}~\bibnamefont {Ellis}},
  \bibinfo {author} {\bibfnamefont {M.}~\bibnamefont {Chiba}}, \bibinfo
  {author} {\bibfnamefont {J.~E.}\ \bibnamefont {Greene}}, \bibinfo {author}
  {\bibfnamefont {H.}~\bibnamefont {Aihara}}, \bibinfo {author} {\bibfnamefont
  {N.}~\bibnamefont {Arimoto}}, \bibinfo {author} {\bibfnamefont
  {K.}~\bibnamefont {Bundy}}, \bibinfo {author} {\bibfnamefont
  {J.}~\bibnamefont {Cohen}}, \bibinfo {author} {\bibfnamefont
  {O.}~\bibnamefont {Dor{\'e}}}, \bibinfo {author} {\bibfnamefont
  {G.}~\bibnamefont {Graves}}, \bibinfo {author} {\bibfnamefont {J.~E.}\
  \bibnamefont {Gunn}}, \bibinfo {author} {\bibfnamefont {T.}~\bibnamefont
  {Heckman}}, \bibinfo {author} {\bibfnamefont {C.}~\bibnamefont {Hirata}},
  \bibinfo {author} {\bibfnamefont {P.}~\bibnamefont {Ho}}, \bibinfo {author}
  {\bibfnamefont {J.-P.}\ \bibnamefont {Kneib}}, \bibinfo {author}
  {\bibfnamefont {O.~L.}\ \bibnamefont {F{\`e}vre}}, \bibinfo {author}
  {\bibfnamefont {L.}~\bibnamefont {Lin}}, \bibinfo {author} {\bibfnamefont
  {S.}~\bibnamefont {More}}, \bibinfo {author} {\bibfnamefont {H.}~\bibnamefont
  {Murayama}}, \bibinfo {author} {\bibfnamefont {T.}~\bibnamefont {Nagao}},
  \bibinfo {author} {\bibfnamefont {M.}~\bibnamefont {Ouchi}}, \bibinfo
  {author} {\bibfnamefont {M.}~\bibnamefont {Seiffert}}, \bibinfo {author}
  {\bibfnamefont {J.}~\bibnamefont {Silverman}}, \bibinfo {author}
  {\bibfnamefont {L.~S.}\ \bibnamefont {Jr}}, \bibinfo {author} {\bibfnamefont
  {D.~N.}\ \bibnamefont {Spergel}}, \bibinfo {author} {\bibfnamefont {M.~A.}\
  \bibnamefont {Strauss}}, \bibinfo {author} {\bibfnamefont {H.}~\bibnamefont
  {Sugai}}, \bibinfo {author} {\bibfnamefont {Y.}~\bibnamefont {Suto}},
  \bibinfo {author} {\bibfnamefont {H.}~\bibnamefont {Takami}}, \ and\ \bibinfo
  {author} {\bibfnamefont {R.}~\bibnamefont {Wyse}},\ }\href
  {http://arxiv.org/abs/1206.0737} {\  (\bibinfo {year} {2012})},\ \Eprint
  {http://arxiv.org/abs/1206.0737} {1206.0737} \BibitemShut {NoStop}%
\bibitem [{\citenamefont {Dor{\'e}}\ \emph {et~al.}(2014)\citenamefont
  {Dor{\'e}}, \citenamefont {Bock}, \citenamefont {Ashby}, \citenamefont
  {Capak}, \citenamefont {Cooray}, \citenamefont {de~Putter}, \citenamefont
  {Eifler}, \citenamefont {Flagey}, \citenamefont {Gong}, \citenamefont
  {Habib}, \citenamefont {Heitmann}, \citenamefont {Hirata}, \citenamefont
  {Jeong}, \citenamefont {Katti}, \citenamefont {Korngut}, \citenamefont
  {Krause}, \citenamefont {Lee}, \citenamefont {Masters}, \citenamefont
  {Mauskopf}, \citenamefont {Melnick}, \citenamefont {Mennesson}, \citenamefont
  {Nguyen}, \citenamefont {{\"O}berg}, \citenamefont {Pullen}, \citenamefont
  {Raccanelli}, \citenamefont {Smith}, \citenamefont {Song}, \citenamefont
  {Tolls}, \citenamefont {Unwin}, \citenamefont {Venumadhav}, \citenamefont
  {Viero}, \citenamefont {Werner},\ and\ \citenamefont {Zemcov}}]{spherex}%
  \BibitemOpen
  \bibfield  {author} {\bibinfo {author} {\bibfnamefont {O.}~\bibnamefont
  {Dor{\'e}}}, \bibinfo {author} {\bibfnamefont {J.}~\bibnamefont {Bock}},
  \bibinfo {author} {\bibfnamefont {M.}~\bibnamefont {Ashby}}, \bibinfo
  {author} {\bibfnamefont {P.}~\bibnamefont {Capak}}, \bibinfo {author}
  {\bibfnamefont {A.}~\bibnamefont {Cooray}}, \bibinfo {author} {\bibfnamefont
  {R.}~\bibnamefont {de~Putter}}, \bibinfo {author} {\bibfnamefont
  {T.}~\bibnamefont {Eifler}}, \bibinfo {author} {\bibfnamefont
  {N.}~\bibnamefont {Flagey}}, \bibinfo {author} {\bibfnamefont
  {Y.}~\bibnamefont {Gong}}, \bibinfo {author} {\bibfnamefont {S.}~\bibnamefont
  {Habib}}, \bibinfo {author} {\bibfnamefont {K.}~\bibnamefont {Heitmann}},
  \bibinfo {author} {\bibfnamefont {C.}~\bibnamefont {Hirata}}, \bibinfo
  {author} {\bibfnamefont {W.-S.}\ \bibnamefont {Jeong}}, \bibinfo {author}
  {\bibfnamefont {R.}~\bibnamefont {Katti}}, \bibinfo {author} {\bibfnamefont
  {P.}~\bibnamefont {Korngut}}, \bibinfo {author} {\bibfnamefont
  {E.}~\bibnamefont {Krause}}, \bibinfo {author} {\bibfnamefont {D.-H.}\
  \bibnamefont {Lee}}, \bibinfo {author} {\bibfnamefont {D.}~\bibnamefont
  {Masters}}, \bibinfo {author} {\bibfnamefont {P.}~\bibnamefont {Mauskopf}},
  \bibinfo {author} {\bibfnamefont {G.}~\bibnamefont {Melnick}}, \bibinfo
  {author} {\bibfnamefont {B.}~\bibnamefont {Mennesson}}, \bibinfo {author}
  {\bibfnamefont {H.}~\bibnamefont {Nguyen}}, \bibinfo {author} {\bibfnamefont
  {K.}~\bibnamefont {{\"O}berg}}, \bibinfo {author} {\bibfnamefont
  {A.}~\bibnamefont {Pullen}}, \bibinfo {author} {\bibfnamefont
  {A.}~\bibnamefont {Raccanelli}}, \bibinfo {author} {\bibfnamefont
  {R.}~\bibnamefont {Smith}}, \bibinfo {author} {\bibfnamefont {Y.-S.}\
  \bibnamefont {Song}}, \bibinfo {author} {\bibfnamefont {V.}~\bibnamefont
  {Tolls}}, \bibinfo {author} {\bibfnamefont {S.}~\bibnamefont {Unwin}},
  \bibinfo {author} {\bibfnamefont {T.}~\bibnamefont {Venumadhav}}, \bibinfo
  {author} {\bibfnamefont {M.}~\bibnamefont {Viero}}, \bibinfo {author}
  {\bibfnamefont {M.}~\bibnamefont {Werner}}, \ and\ \bibinfo {author}
  {\bibfnamefont {M.}~\bibnamefont {Zemcov}},\ }\href
  {http://arxiv.org/abs/1412.4872} {\  (\bibinfo {year} {2014})},\ \Eprint
  {http://arxiv.org/abs/1412.4872} {1412.4872} \BibitemShut {NoStop}%
\bibitem [{\citenamefont {Cutler}\ and\ \citenamefont
  {Holz}(2009)}]{Cutler:2009}%
  \BibitemOpen
  \bibfield  {author} {\bibinfo {author} {\bibfnamefont {C.}~\bibnamefont
  {Cutler}}\ and\ \bibinfo {author} {\bibfnamefont {D.~E.}\ \bibnamefont
  {Holz}},\ }\href {http://arxiv.org/abs/0906.3752} {\  (\bibinfo {year}
  {2009})},\ \Eprint {http://arxiv.org/abs/0906.3752} {0906.3752} \BibitemShut
  {NoStop}%
\bibitem [{\citenamefont {Laguna}\ \emph {et~al.}(2009)\citenamefont {Laguna},
  \citenamefont {Larson}, \citenamefont {Spergel},\ and\ \citenamefont
  {Yunes}}]{Laguna:2009}%
  \BibitemOpen
  \bibfield  {author} {\bibinfo {author} {\bibfnamefont {P.}~\bibnamefont
  {Laguna}}, \bibinfo {author} {\bibfnamefont {S.~L.}\ \bibnamefont {Larson}},
  \bibinfo {author} {\bibfnamefont {D.}~\bibnamefont {Spergel}}, \ and\
  \bibinfo {author} {\bibfnamefont {N.}~\bibnamefont {Yunes}},\ }\href
  {http://arxiv.org/abs/0905.1908} {\  (\bibinfo {year} {2009})},\ \Eprint
  {http://arxiv.org/abs/0905.1908} {0905.1908} \BibitemShut {NoStop}%
\bibitem [{\citenamefont {Collett}\ and\ \citenamefont
  {Bacon}(2016)}]{Collett:2016}%
  \BibitemOpen
  \bibfield  {author} {\bibinfo {author} {\bibfnamefont {T.~E.}\ \bibnamefont
  {Collett}}\ and\ \bibinfo {author} {\bibfnamefont {D.}~\bibnamefont
  {Bacon}},\ }\href {http://arxiv.org/abs/1602.05882} {\  (\bibinfo {year}
  {2016})},\ \Eprint {http://arxiv.org/abs/1602.05882} {1602.05882}
  \BibitemShut {NoStop}%
\end{thebibliography}%

\label{lastpage}
\end{document}